\documentclass[twocolumn]{aastex631}
\usepackage{natbib}
\usepackage{amsmath}

\usepackage{color}
\newcommand{\fdir}{./}
\newcommand{\teff}{\log(T_{\rm eff}/\mbox{K})}

\newcommand{\xeff}{\chi_{\rm eff}}

\newcommand{\msun}{M_\odot}

\newcommand{\rsun}{R_\odot}
\newcommand{\zsun}{Z_\odot}
\newcommand{\zbar}{\overline{Z}}
\newcommand{\fb}{f_{\rm b}}
\newcommand{\kms}{{\rm km~s}^{-1}}
\newcommand{\cago}{^{12}{\rm C}(\alpha,\gamma)^{16}{\rm O}}
\newcommand{\tsig}{3\sigma}
\newcommand{\rate}{{\cal R}}
\newcommand{\prop}{{\cal P}}
\newcommand{\neff}{{\cal N}}
\newcommand{\nint}{\eta}
\newcommand{\thb}{t_{\rm h}}
\newcommand{\tlb}{t_{\rm lb}}
\newcommand{\td}{t_{\rm d}}
\newcommand{\sigmaz}{\sigma_{\rm Z}}
\newcommand{\erase}{\bgroup\markoverwith{\rule[.5ex]{2pt}{0.4pt}}\ULon}

\begin{document}

\title{Merger rate density of binary black holes through isolated
  Population I, II, III and extremely metal-poor binary star
  evolution}

\correspondingauthor{Ataru Tanikawa}
\email{tanikawa@ea.c.u-tokyo.ac.jp}

\author{Ataru Tanikawa}
\affiliation{Department of Earth Science and Astronomy, College of
  Arts and Sciences, The University of Tokyo, 3-8-1 Komaba, Meguro-ku,
  Tokyo 153-8902, Japan}

\author{Takashi Yoshida}
\affiliation{Yukawa Institute for Theoretical Physics, Kyoto
  University, Kitashirakawa Oiwakecho, Sakyo-ku, Kyoto 606-8502,
  Japan}

\author{Tomoya Kinugawa}
\affiliation{Institute for Cosmic Ray Research, The University of
  Tokyo, Kashiwa, Chiba}

\author{Alessandro A. Trani}
\affiliation{Department of Earth Science and Astronomy, College of
  Arts and Sciences, The University of Tokyo, 3-8-1 Komaba, Meguro-ku,
  Tokyo 153-8902, Japan}
\affiliation{Okinawa Institute of Science and Technology, Quantum
  Gravity unit, 1919-1 Tancha, Onna-son, Okinawa 904-0495, Japan}

\author{Takashi Hosokawa}
\affiliation{Department of Physics, Graduate School of Science, Kyoto
  University, Sakyo, Kyoto 606-8502, Japan}

\author{Hajime Susa}
\affiliation{Department of Physics, Faculty of Science and Engineering, Konan University, Kobe 658-8501, Japan}

\author{Kazuyuki Omukai}
\affiliation{Astronomical Institute, Graduate School of Science,
  Tohoku University, Aoba, Sendai 980-8578, Japan}

\begin{abstract}
  We investigate the formation of merging binary black holes (BHs)
  through isolated binary evolution, performing binary population
  synthesis calculations covering an unprecedentedly wide metallicity
  range of Population (Pop) I, II, III, and extremely metal-poor (EMP)
  binary stars.  We find that the predicted merger rate density and
  primary BH mass ($m_1$) distribution are consistent with the
  gravitational wave (GW) observations. Notably, Pop III and EMP ($<
  10^{-2}$ $Z_\odot$) binary stars yield most of the pair instability
  (PI) mass gap events with $m_1 = 65$--$130$ $M_\odot$. Pop III
  binary stars contribute more to the PI mass gap events with
  increasing redshift, and all the PI mass gap events have the Pop III
  origin at redshifts $\gtrsim 8$. Our result can be assessed by
  future GW observations in the following two points. First, there are
  no binary BHs with $m_1=100$--$130$ $M_\odot$ in our result, and
  thus the $m_1$ distribution should suddenly drop in the range of
  $m_1=100$--$130$ $M_\odot$. Second, the PI mass gap event rate
  should increase toward higher redshift up to $\sim 11$, since those
  events mainly originate from the Pop III binary stars. We find that
  the following three assumptions are needed to reproduce the current
  GW observations: a top-heavy stellar initial mass function and the
  presence of close binary stars for Pop III and EMP binary stars, and
  inefficient convective overshoot in the main-sequence phase of
  stellar evolution. Without any of the above, the number of PI mass
  gap events becomes too low to reproduce current GW observations.
\end{abstract}


\section{Introduction}
\label{sec:Introduction}

A growing number of binary black hole (BH) mergers have been
discovered by gravitational wave (GW) observatories, reaching nearly
$50$ by the end of the first half of the third observing run
\citep{2019PhRvX...9c1040A, 2021PhRvX..11b1053A, 2021arXiv210801045T}.
The number of binary BHs has grown to about $90$
\citep{2021arXiv211103606T}, while this paper is under review. The
origin of merging binary BHs is still an open question \citep[see
  review by][]{2018arXiv180605820M, 2021NatAs...5..749G,
  2021hgwa.bookE...4M}. The progenitors can be isolated Population
(Pop) I/II binary stars that go through common envelope evolution
\citep{1998ApJ...506..780B, 2016Natur.534..512B, 2020A&A...636A.104B,
  2017NatCo...814906S, 2017PASA...34...58E, 2019MNRAS.482..870E,
  2017MNRAS.472.2422M, 2019MNRAS.487....2M, 2018MNRAS.481.1908K,
  2020ApJ...901L..39O}, through chemically homogeneous evolution
\citep{2016MNRAS.458.2634M, 2016A&A...588A..50M, 2021MNRAS.505..663R},
and through stable mass transfer \citep{2017MNRAS.471.4256V,
  2019MNRAS.490.3740N, 2021A&A...647A.153B}. Isolated Pop III binary
stars are also suggested to form binary BHs mainly through stable mass
transfer \citep{2014MNRAS.442.2963K, 2016MNRAS.456.1093K,
  2017MNRAS.468.5020I}, and partly through common envelope evolution
\citep{2020MNRAS.498.3946K, 2021MNRAS.501L..49K, 2021PTEP.2021b1E01K,
  2021MNRAS.505.2170T, 2021ApJ...910...30T}.  Merging binary BHs can
be formed through dynamical capture in globular clusters
\citep{2000ApJ...528L..17P, 2006ApJ...637..937O, 2008ApJ...676.1162S,
  2010MNRAS.407.1946D, 2011MNRAS.416..133D, 2013MNRAS.435.1358T,
  2014MNRAS.440.2714B, 2016PhRvD..93h4029R, 2016ApJ...824L...8R,
  2018PhRvL.120o1101R,2019PhRvD.100d3027R, 2017MNRAS.464L..36A,
  2017PASJ...69...94F, 2017MNRAS.469.4665P, 2018ApJ...855..124S,
  2018PhRvD..97j3014S, 2018MNRAS.480.5645H, 2020MNRAS.498.4287H,
  2020PASA...37...44A, 2020ApJ...903...45K, 2021ApJ...907L..25W,
  2021MNRAS.504..910T, 2021MNRAS.504.5778W}, in young star clusters
\citep{2010MNRAS.402..371B, 2017MNRAS.467..524B, 2018MNRAS.473..909B,
  2018MNRAS.481.5123B, 2021MNRAS.500.3002B, 2021MNRAS.503.3371B,
  2014MNRAS.441.3703Z, 2016MNRAS.459.3432M, 2019MNRAS.483.1233R,
  2021MNRAS.507.3612R, 2019MNRAS.486.3942K, 2020MNRAS.495.4268K,
  2021arXiv210209323K, 2019MNRAS.487.2947D, 2020MNRAS.497.1043D,
  2020MNRAS.498..495D, 2020ApJ...898..152S, 2021MNRAS.502.4877S,
  2020ApJ...902L..26F, 2021ApJ...913L..29F, 2021arXiv210507003A}, at
galactic centers \citep{2009MNRAS.395.2127O, 2016ApJ...831..187A,
  2017MNRAS.464..946S, 2018MNRAS.474.5672L, 2018ApJ...866...66M,
  2019ApJ...881...20R, 2019ApJ...885..135T, 2019ApJ...876..122Y,
  2020ApJ...891...47A, 2021Symm...13.1678M, 2021MNRAS.505..339M,
  2020ApJ...899...26T, 2020ApJ...898...25T, 2021ApJ...907L..20T,
  2021ApJ...908..194T}, and/or in Pop III star clusters
\citep{2020MNRAS.495.2475L, 2020ApJ...903L..40L, 2021MNRAS.506.5451L,
  2021MNRAS.501..643L}. BH mergers can be also triggered by secular
evolution in hierarchical multiple stellar systems
\citep{2014ApJ...781...45A, 2017ApJ...841...77A, 2017ApJ...836...39S,
  2018ApJ...863....7R, 2019MNRAS.486.4781F, 2020ApJ...895L..15F,
  2020ApJ...898...99H, 2021MNRAS.506.5345H, 2021arXiv211014680V}, and
around massive BHs at galactic centers \citep{2012ApJ...757...27A,
  2016ApJ...828...77V, 2018ApJ...865....2H,
  2018ApJ...856..140H}. Merging binary BHs can be primordial origin
\citep{2021arXiv210503349F}. Multiple origins have been suggested
\citep{2021MNRAS.507.5224B, 2021ApJ...910..152Z, 2021ApJ...913L...5N,
  2021JCAP...03..068H}.

The isolated binary scenario, however, has been challenged by the
discovery of GW190521 with a $\sim 90$ $\msun$ BH with the $90$ \%
credible interval \citep{2020PhRvL.125j1102A}, which is prohibited to
form by the pulsational pair instability (PPI) and pair instability
(PI) supernovae \citep{2003ApJ...591..288H, 2016A&A...594A..97B,
  2017MNRAS.470.4739S, 2017ApJ...836..244W, 2019ApJ...878...49W,
  2018MNRAS.474.2959G}, in contrast to the hierarchical merger
scenario in star clusters \citep{2019PhRvD.100d3027R,
  2020MNRAS.497.1043D, 2020ApJ...903L..40L, 2020arXiv201006161A,
  2021ApJ...908..194T, 2021ApJ...915L..35K, 2021MNRAS.505..339M,
  2021MNRAS.502.2049L, 2021arXiv210410253G, 2021NatAs...5..749G,
  2021arXiv210507003A}. There are, however, still some way outs to
accommodate the existence of GW190521 in the framework of the isolated
binary scenario, which can be classified into three categories. The
first possibility is that GW190521 actually straddles the PI mass gap;
the primary and secondary BHs have masses beyond and below the PI mass
gap, respectively \citep{2020ApJ...904L..26F, 2021ApJ...907L...9N,
  2021arXiv210506360E}. The second is that the PI mass gap may shift
upward because of microphysics in single star evolution different from
what has been thought: e.g., $\cago$ reaction rate three times smaller
than the standard one \citep{2019ApJ...887...53F, 2020ApJ...902L..36F,
  2021MNRAS.501.4514C}, or dark matter effects
\citep{2020arXiv200707889C, 2021PhRvD.104d3015Z}. In particular,
\cite{2020ApJ...905L..15B} have shown that isolated binary stars can
form the PI mass gap events consistent with GW190521-like event rate,
taking into account the reduced $\cago$ reaction rate. The third is
that the lower PI mass gap, $\sim 65$--$100$ $\msun$, may disappear
with more detailed consideration of macrophysics in single star
evolution, such as metal-poor star evolution
\citep{2021MNRAS.501L..49K, 2021MNRAS.502L..40F, 2021MNRAS.504..146V,
  2021MNRAS.505.2170T}, strength of stellar wind mass loss
\citep{2020ApJ...890..113B}, and stellar rotation effect
\citep{2020ApJ...888...76M}. In this paper, we focus on the third
possibility, to be more precise metal-poor star evolution, to
accommodate the formation of GW190521-like events in the isolated
binary scenario.

\cite{2021MNRAS.505.2170T} have shown that Pop III stars can produce
sufficient PI mass gap events to be reconciled with the GW190521-like
event rate inferred by GW observations, if convective overshoot
between their convective core and radiative envelope is inefficient
during the main-sequence (MS) phase. Pop III stars, however, produce
fewer binary BH mergers in the lower mass range of $\lesssim 50$
$\msun$ than the expectation from GW observations. Thus, we need to
take into account not only Pop III stars but also Pop I/II stars to
explain all the observed BH mergers in the framework of the isolated
binary scenario.

For this purpose, we perform binary population synthesis (BPS)
calculations for isolated Pop I/II/III and extremely metal-poor (EMP)
binary stars in an unprecedentedly wide metallicity range in this
paper. The reason for remarking the ``unprecedentedly wide metallicity
range'' is that we cover not only Pop I/II stars but also EMP and Pop
III stars, which are strictly defined at the end of this
section. Previous studies of binary BHs have focused on only Pop I/II
stars as most of BPS codes support only Pop I/II stars like SeBa
\citep{1996A&A...309..179P, 2012A&A...546A..70T}, BSE
\citep{2002MNRAS.329..897H}, binary\_c \citep{2009A&A...508.1359I},
MOBSE \citep{2018MNRAS.474.2959G}, COSMIC \citep{2020ApJ...898...71B},
and COMPAS \citep{2021arXiv210910352T}. Some studies have examined Pop
III stars \citep{2014MNRAS.442.2963K, 2016MNRAS.460L..74H}, and
StarTrack \citep{2002ApJ...572..407B, 2017MNRAS.471.4702B} can cover
Pop I/II/III stars. However, no studies have included EMP stars.

We investigate isolated Pop I/II binary evolution by our own
calculations rather than making use of the results of previous studies
on isolated Pop I/II binary stars as done by
\cite{2021MNRAS.504L..28K}. We do this for consistency. We construct
our own Pop III star evolution model as in \cite{2021MNRAS.505.2170T},
decreasing the metallicity of our own Pop I/II star evolution model,
rather than adopting the widely-used SSE Pop I/II star evolution model
\citep{2000MNRAS.315..543H}.  It is unclear if our own Pop I/II star
evolution models yield similar binary BHs to the SSE Pop I/II star
evolution models, or if our own model Pop I/II star evolution models
can produce binary BHs in the mass range of $\lesssim 50$ $\msun$
consistently with the GW observations, and do not overproduce the PI
mass gap events compared to the GW observations. We expect that the
SSE and our own Pop I star evolution models would yield similar binary
BHs, since these models are similar because of similar calibration to
nearby Pop I stars including the Sun \citep{1998MNRAS.298..525P,
  2000MNRAS.315..543H, Yoshida19}. On the other hand, this may not be
the case for the SSE and our own Pop II evolution models. Therefore,
we use the SSE model for Pop I stars, and our own model for Pop II
stars by our own calculations. We could have borrowed Pop I binary BHs
from previous studies. However, we do not so. This enables us to
examine Pop I binary BHs as well as Pop II, III and EMP binary BHs in
more detail, since we obtain their properties by ourselves. For
example, we can investigate the redshift evolution of binary BH
mergers formed from Pop I/II/III and EMP binary stars.

In this paper, we examine the isolated binary scenario by the BPS
calculations, and demonstrate that from the isolated Pop I/II/III and
EMP binary stars the observed BH merger rate density and its primary
BH mass distribution over $\sim 5$--$100$ $\msun$ are successfully
reproduced under the assumption of inefficient convective overshoot.
This means that isolated Pop III (and EMP) binaries can form
GW190521-like events, and that isolated Pop I/II star evolution models
can produce binary BHs in the mass range of $\lesssim 50$ $\msun$
consistently with GW observations without overproducing PI mass gap
events regardless of the choice of Pop I/II star evolution models. We
also predict the redshift evolution of merging binary BHs, which
should help assessing the validity of our model by future GW
observations, e.g., by Cosmic Explorer \citep{2019BAAS...51g..35R} and
Einstein Telescope \citep{2010CQGra..27s4002P}. We also point out the
key assumptions allowing the predicted BH properties to be consistent
with the observations, considering large uncertainties in the initial
conditions and evolution of Pop III stars due to its non detection so
far \citep{2015ARA&A..53..631F, 2018MNRAS.473.5308M,
  2019MNRAS.487..486M}.

The structure of this paper is as follows. We present our numerical
method to calculate the BH merger rate density at a given redshift in
section \ref{sec:Method}. We present and discuss the numerical results
in section \ref{sec:Results} and \ref{sec:Discussion},
respectively. We summarize this paper in section \ref{sec:Summary}. We
adopt the cosmological parameters based on the latest observation by
\cite{2020A&A...641A...6P}. We define metallicities of Pop I, Pop II,
EMP, and Pop III stars as $Z/\zsun > 0.16$, $10^{-3} < Z/\zsun \le
0.16$, $0 < Z/\zsun \le 10^{-3}$, and $Z/\zsun=0$, respectively, where
$Z$ is the metallicity, $\zsun$ is the solar metallicity set to $\zsun
= 0.02$. We treat $Z/\zsun = 10^{-8}$ stars as Pop III stars,
$Z/\zsun=10^{-6}$ and $10^{-4}$ as EMP stars, $Z/\zsun=10^{-2}, 0.025,
0.05$ and $0.1$ as Pop II stars, and $Z/\zsun=0.25, 0.5$ and $1$ as
Pop I stars.

\section{Method}
\label{sec:Method}

We describe initial conditions for our BPS calculations in section
\ref{sec:InitialConditions}. We overview our single and binary star
models for our BPS calculations in section
\ref{sec:SingleAndBinaryStarModels}. As we take into account several
different initial conditions and single/binary star models, we
describe our parameter sets for them in section
\ref{sec:ParameterSets}. In section
\ref{sec:DifferentialBhMergerRateDensity}, we formulate how to obtain
the BH merger rate density and its derivatives, using our BPS
calculation results.

\subsection{Initial conditions}
\label{sec:InitialConditions}

We account for the star formation rate (SFR) density with two
components as
\begin{align}
  \phi(z) = \phi_{\rm I/II}(z) + \phi_{\rm III}(z), \label{eq:TotalSfr}
\end{align}
where $z$ is the redshift, and $\phi_{\rm I/II}(z)$ and $\phi_{\rm
  III}(z)$ are the Pop I/II and Pop III SFR densities, respectively. We adopt
the Pop I/II SFR density of \cite{2017ApJ...840...39M}, expressed as
\begin{align}
  \phi_{\rm I/II} (z) = 0.01 \frac{(1+z)^{2.6}}{1+\left[ \left( 1 + z
      \right) / 3.2 \right]^{6.2}} \; \msun \; {\rm yr}^{-1} \; {\rm
    Mpc}^{-3}.
\end{align}
We assume the metallicity distribution of Pop I/II components at a
given $z$ as a logarithmic normal distribution similarly to
\cite{2020A&A...636A.104B} and \cite{2021MNRAS.502.4877S}, given by
\begin{align}
  p_{\rm I/II}(z, Z) &= \frac{1}{(2 \pi \sigmaz^2)^{1/2}} \nonumber \\
  &\times \exp \left\{ - \frac{ \left[ \log (Z/\zsun) - \log
      (\zbar_{\rm I/II}(z)/\zsun) \right]^2 }{2 \sigmaz^2} \right\},
\end{align}
where $\zbar_{\rm I/II}(z)$ is the average metallicity of the Pop I/II
component at a given redshift $z$, and $\sigmaz$ is the metallicity
dispersion set to $\sigmaz = 0.35$. The average metallicity evolution
in the Pop I/II component $\zbar_{\rm I/II}(z)$ is considered in the
same way as \cite{2017ApJ...840...39M}:
\begin{align}
  \log \left( \zbar_{\rm I/II}(z) / \zsun \right) = 0.153 - 0.074
  z^{1.34}. \label{eq:metalEvol}
\end{align}
We use the following Pop III SFR density, obtained by simplifying the
simulation result of \cite{2020MNRAS.492.4386S} as shown in their
figure 4:
\begin{align}
  &\log ( \phi_{\rm III} (t) / \msun \; {\rm yr}^{-1} \; {\rm
    Mpc}^{-3} ) = \nonumber \\
  &\left\{
  \begin{array}{ll}
    0.0130   (t/{\rm Myr} - 100) - 5.00 & (100 \le t/{\rm Myr} < 200) \\
    -0.00261 (t/{\rm Myr} - 200) - 3.70 & (200 \le t/{\rm Myr} < 400) \\
    -0.0108  (t/{\rm Myr} - 400) - 4.22 & (400 \le t/{\rm Myr} < 500) \\
    0                                 & (\mbox{otherwise})
  \end{array}
  \right., \label{eq:PopIIISfr}
\end{align}
where $t$ indicates the cosmic time since the Big Bang. The three
formulae in Eq. (\ref{eq:PopIIISfr}) are not continuous, since the
numbers in the three formulae are written to $3$ significant
digits. We regard that $Z=10^{-8} \zsun$ stars, which are the most
metal-poor in our single star model, are in our simulations equivalent
to Pop III (i.e. zero-metal) stars. This is because their evolution is
almost identical. Thus, we suppose the metallicity distribution of the
Pop III component as
\begin{align}
  p_{\rm III}(z,Z) &= \delta (Z-10^{-8}\zsun) \\
  &= \left\{
  \begin{array}{ll}
    \infty & (Z/\zsun = 10^{-8}) \\
    0      & ({\rm otherwise})
  \end{array}
  \right.,
\end{align}
where the $\delta$ function has the usual property as seen above. This
means that $Z=10^{-8} \zsun$ (i.e. Pop III or zero-metal) stars have
the SFR density described in Eq. (\ref{eq:PopIIISfr}) throughout all
the times (or redshifts).  Then, we can get the SFR density
differentiated by the metallicity:
\begin{align}
  \frac{d\phi(z)}{dZ} = \phi_{\rm I/II}(z) p_{\rm I/II}(z,Z) +
  \phi_{\rm III}(z)p_{\rm III}(z,Z). \label{eq:SfrDiffZ}
\end{align}
This equation is used in section
\ref{sec:DifferentialBhMergerRateDensity}.

We assume that the binary number fraction is $0.5$ for all
metallicities, where the fraction is close to an intrinsic binary
fraction $0.69^{+0.09}_{-0.09}$ found by
\cite{2012Sci...337..444S}. The initial mass function (IMF) of the
single stars and primary stars of the binary stars is given by a
combination of the following two mass functions. The first is the
Kroupa's IMF \citep{2001MNRAS.322..231K} written as
\begin{align}
  f(m) dm \propto \left\{ 
  \begin{array}{ll}
    m^{-1.3} & (0.08 \le m/\msun <   0.5) \\
    m^{-2.3} & (0.5  \le m/\msun \le 150) \\
  \end{array}
\right.,
\end{align}
where $m$ is the stellar mass. The second is a function flat in the logarithmic
scale (hereafter, the logarithmically flat IMF), such that
\begin{align}
  f(m) dm \propto m^{-1} & (10 \le m/\msun \le 150).
\end{align}

We add the above two mass functions with a weight such that more stars
follow the logarithmically flat IMF at lower metallicities. We adopt
three different methods to weight the two functions. The first is to
adopt just the Kroupa's IMF for all the metallicities. We call this
IMF model the ``top-light IMF model''. The second is to use
the Kroupa's IMF for $Z/\zsun \ge 10^{-5}$ and otherwise the
logarithmically flat IMF. This is because the stellar IMF is
theoretically predicted to transition from top-light to top-heavy at
$Z/\zsun \sim 10^{-5}$ \citep{2004ARA&A..42...79B,
  2005ApJ...626..627O, 2006MNRAS.369..825S, 2010MNRAS.407.1003M}. We
call this IMF model the ``stepwise IMF model''. In the third method,
the IMF shifts gradually from Kroupa's IMF to the logarithmically flat
one with decreasing metallicity. All the stars follow the Kroupa's IMF
for $Z/\zsun > 10^{-2}$. At $Z/\zsun =10^{-2}$, $60$ \% of them obey
the logarithmically flat IMFs in terms of mass, while this fraction
increases to $90$ \% at $Z/\zsun = 10^{-4}$. At metallicity as low as
$Z/\zsun \le 10^{-6}$, all the stars follow the logarithmically flat
distribution. This behavior is inspired from the simulation results of
\cite{2021arXiv210304997C}. We call this IMF model the ``transitional
IMF model''.

For the initial distributions of binary parameters we refer to
\cite{2012Sci...337..444S}, who proposed the following expressions for
the mass ratios ($q$), periods ($P$), and orbital eccentricities ($e$)
of binary stars:
\begin{align}
  &f(q) \propto q^{-0.1} \; (0.1 \le q \le 1) \label{eq:fq} \\
  &f(\log P) \propto (\log P)^{-0.55} \; (0.15 < \log (P/{\rm day}) <
  5.5) \label{eq:fP} \\
  &f(e) \propto e^{-0.42} \; (0 \le e \le 1), \label{eq:fe}
\end{align}
where $q$ is defined as the ratio of the secondary mass to the primary
mass. We call this binary star initial condition the ``Sana's
model''. The period range in Eq. (\ref{eq:fP}) is the same as one
adopted by \cite{2015ApJ...814...58D}. On the other hand, our binary
number fraction ($0.5$) is different from that of
\cite{2015ApJ...814...58D} who have adopted $1$ for the binary number
fraction as default by careful consideration of the difference between
their and Sana's period ranges.

In addition to this, we introduce another binary star initial
condition, where we exclude binary stars with pericenter distances of
less than $200$ $\rsun$ when $Z \le 10^{-2} \zsun$. This condition is
inspired by the following consideration. $Z/\zsun \lesssim 10^{-2}$
stars tend to expand to $\sim 100$ $\rsun$ in radius during their
protostellar evolution thanks to the high mass accretion rate
\citep{2009ApJ...691..823H, 2009ApJ...703.1810H, 2010ApJ...722.1793O}.
Even if a metal-poor star initially has a close companion star, they
may merge during their protostar phase, and may not make a close
binary system. We call this binary initial condition the ``wide binary
model''. Nonetheless, we still adopt the Sana's model as the fiducial
one.  This is because the radial expansion depends on the geometry of
protostellar accretion, which is uncertain. If the accretion shifts
from spherical to disk accretion relatively early on, the maximum
expansion radius would be less than $100$ $\rsun$
\citep{2010ApJ...721..478H}.  In addition, even with $Z < 10^{-2}
\zsun$ close binaries can be formed through $N$-body interaction after
their maximal expansion phase \citep[e.g.][]{2021MNRAS.501..643L}.

\subsection{Single and binary star models}
\label{sec:SingleAndBinaryStarModels}

We employ the SSE model for single stars with $5 \times 10^{-3} \le
Z/\zsun \le 1.5$ \citep{2000MNRAS.315..543H}, and M and L
models\footnote{The two models are named after the {\bf M}ilky way and
  the {\bf L}arge Magellanic Cloud, whose stars were used to calibrate
  the overshoot parameter of the M and L models, respectively
  \citep{Yoshida19}.} for single stars with $Z/\zsun = 10^{-8},
10^{-6}, 10^{-4}$, and $10^{-2} \le Z/\zsun \le 0.1$
\citep{2020MNRAS.495.4170T, 2021MNRAS.505.2170T}.  An L-model star
experiences more efficient convective overshoot than an M-model star
does, for which the convective overshoot happens between the
convective core and radiative envelope in the MS phase. The L-model
star has a larger He core than the M-model star at the end of their MS
phase if these stars have the same initial mass. The L-model star
expands much more in radius than the M-model star at their post-MS
phase. We summarize the details of the M/L models, and compare them
with the SSE models in appendix \ref{sec:SingleStarModels}. We
describe later in this section how to adopt the SSE, M, and L models.

We take into account stellar wind mass-loss in the same way as
\cite{2020MNRAS.495.4170T}, based on the wind model of
\cite{2010ApJ...714.1217B} with modification of metallicity-dependent
luminous blue variable winds \citep{2018MNRAS.474.2959G}.  For
simplicity, we do not include rotational enhancement of stellar winds
\cite[unlike as in][]{2020MNRAS.495.4170T}.

Our supernova model is the rapid model \citep{2012ApJ...749...91F}
with the PI mass loss. The PI mass loss modifies a remnant mass given
by the rapid model ($m_{\rm rapid}$) as follows:
\begin{align}
  m_{\rm rem} = \left\{
  \begin{array}{ll}
    m_{\rm rapid}    & (m_{\rm c,He} \le m_{\rm c,He,PPI}) \\
    m_{\rm c,He,PPI} & (m_{\rm c,He,PPI} < m_{\rm c,He} \le m_{\rm c,He,PISN}) \\
    0                & (m_{\rm c,He,PISN} < m_{\rm c,He} \le m_{\rm c,He,DC}) \\
    m_{\rm rapid}    & (m_{\rm c,He} > m_{\rm c,He,DC}) \\
  \end{array}
  \right., \label{eq:PImodel}
\end{align}
where $m_{\rm rem}$ is the final remnant mass, $m_{\rm c,He}$ is the
He core mass, and $m_{\rm c,He,PPI}$, $m_{\rm c,He,PISN}$, and $m_{\rm
  c,He,DC}$ are the lower He core mass limits of pulsational PI
\citep[PPI:][]{2002ApJ...567..532H, 2007Natur.450..390W,
  2016MNRAS.457..351Y, 2019ApJ...887...72L}, PI supernova
\citep[PISN:][]{1967PhRvL..18..379B, 1968Ap&SS...2...96F,
  1984ApJ...280..825B, 1986A&A...167..274E, 2001ApJ...550..372F,
  2002ApJ...567..532H, 2002ApJ...565..385U}, and direct collapse (DC),
respectively. Note that, although $m_{\rm c,He,PPI}$ is fixed
  to a constant value in the above equation, in reality it should depend on
  $m_{\rm c,He}$ in some complex way \citep[see figure 1
    in][]{2019ApJ...882..121S}. We consider two sets of the mass
limits, since the mass limits strongly depend on the uncertain $\cago$
reaction rate \citep{2020ApJ...902L..36F, 2021MNRAS.501.4514C}. The
first and second sets are $(m_{\rm c,He,PPI}, m_{\rm c,He,PISN},
m_{\rm c,He,DC}) = (45\msun, 65\msun, 135\msun)$ and $(90\msun,
90\msun, 180\msun)$, which are similar to \cite{2016A&A...594A..97B}
and \cite{2020ApJ...905L..15B}, respectively. The first set is based
on the standard $\cago$ reaction rate with respect to the {\tt
  STARLIB} \citep{2013ApJS..207...18S}. Thus, we call the PI model
``the standard PI model''. On the other hand, the second set can be
obtained when the $\cago$ reaction rate is lower than the standard one
by $3 \sigma$. Thus, we call the PI model ``the $\tsig$ PI model''.

\begin{deluxetable*}{c|cccc}
  \tablecaption{Summary of combinations of single star models and PI
    models. \label{tab:SingleStarAndPi}}
  \tablecolumns{5}
  \tablehead{
    Model & Pop III+EMP & Pop II & Pop I & PI \\
          & ($10^{-8},10^{-6},10^{-4}\zsun$) & ($10^{-2},0.025,0.05,0.1\zsun$) & ($0.25,0.5,1\zsun$) &
  }
\startdata
M and SSE-Pop I with standard PI  & M   & M & SSE & Standard \\
L and SSE-Pop I with standard PI  & L   & L & SSE & Standard \\
L and SSE-Pop I with  $\tsig$ PI  & L   & L & SSE & $\tsig$  \\
M and SSE-Pop I/II with standard PI & M & SSE & SSE & Standard  \\
\enddata
\tablecomments{Metallicities in parentheses indicate ones that we
  adopt for our BPS calculations to realize Pop I, Pop II, EMP and Pop
  III stars. We regard $10^{-8}\zsun$ as $0\zsun$.}
\end{deluxetable*}

We have four different combinations of single star models and PI
models as seen in Table \ref{tab:SingleStarAndPi}. In all the
combinations, we adopt the SSE model for Pop I binary stars.  In the
first combination, we adopt the M model for Pop II and Pop III+EMP
stars, and the standard PI model. In the second and third
combinations, we adopt the L model for Pop II and Pop III+EMP stars,
and the standard and $\tsig$ PI models, respectively. In the fourth
combination, we adopt the SSE and M models for Pop II and Pop III+EMP
stars, respectively, and the standard PI model. We do not consider the
$\tsig$ PI model when we adopt the M model for Pop III+EMP stars. This
is because we need either the $\tsig$ PI model or the M model in order
to yield the PI mass gap event, but not both.

We explain how to choose the SSE or M/L models for Pop I, Pop II, EMP,
and Pop III stars as seen in Table \ref{tab:SingleStarAndPi}. We
choose the SSE model for Pop I stars. The practical reason for this
choice is that $Z/\zsun > 0.1$ star evolution models of the M/L models
are not implemented in our BPS code, although these models are
constructed in \cite{Yoshida19}. However, we expect that our results
do not depend on this choice. If we used the M/L models for Pop I
stars in our BPS calculations, these models would be similar to the
SSE model. This is because the M/L models for Pop I stars are based on
the results of \cite{Yoshida19} which are calibrated to nearby Pop I
stars including the Sun, similarly to the SSE model
\citep{1998MNRAS.298..525P, 2000MNRAS.315..543H}. Rather, there is a
possibility that our results are affected by the choice of the SSE and
M/L models for more metal-poor stars, i.e. Pop II stars. However, we
can see that there is little difference yielded by the choice of the
SSE and M/L models for Pop II stars as seen in appendix
\ref{sec:ComparisonWithSse}. Despite that the SSE model supports
metallicity down to $5 \times 10^{-3} \zsun$, we do not use the SSE
model with less than $10^{-2}\zsun$.  The practical reason for this is
that we would like to adopt the same metallicities for Pop II stars
among the SSE and M/L models, and that the M/L models do not support
$5 \times 10^{-3} \le Z/\zsun < 10^{-2}$ star evolution. However, we
can see that the not using the SSE models for this range has no impact
(see details in appendix \ref{sec:EffectsOfBinning}). We adopt the M/L
models for Pop III+EMP stars, since the SSE model does not fully
support Pop III+EMP star evolution. Especially, we regard the M/L
models with $Z/\zsun = 10^{-8}$ as Pop III stars.

We model the BH spin angular momentum ($J_{\rm rem}$) as follows. We
know spin angular momenta of He core ($J_{\rm c,0}$) and H envelope
($J_{\rm e,0}$) for each star at the pre-supernova phase from the
stellar evolution model. We set 
\begin{align}
  J_{\rm rem} = J_{\rm c,0}(m_{\rm c,1}/m_{\rm c,0})+J_{\rm
    e,0}(m_{\rm e,1}/m_{\rm e,0}), \label{eq:CollapsingBhSpin}
\end{align}
where $m_{\rm c,0}$ and $m_{\rm e,0}$ are the masses of He core and H
envelope at the pre-supernova phase, and $m_{\rm c,1}$ and $m_{\rm
  e,1}$ are the masses of He core and H envelope contributing to the
BH mass, i.e., the sum of $m_{\rm c,1}$ and $m_{\rm e,1}$ is the BH
mass. We obtain the above relation, assuming that the He core and H
envelope are in rigid rotation, and they each have uniform
densities. In this prescription, we may underestimate the BH spin
angular momentum, since stellar rotation and density become slower and
smaller outward, respectively. We can calculate the dimensionless BH
spin as
\begin{align}
  \chi_{\rm rem} = \min \left( \frac{J_{\rm rem}}{Gm_{\rm rem}^2/c}, 1
  \right), \label{eq:BhSpin}
\end{align}
where $G$ is the gravitational constant, and $c$ is the light speed.

We include natal kicks of neutron stars and BHs because of asymmetric
supernova explosions. The velocity distribution is given by a single
Maxwellian with $265$ $\kms$ \citep{2005MNRAS.360..974H} if the
remnant acquires no fallback mass. The natal kick velocity is reduced
by $1-\fb$, where $\fb$ is the fraction of the fallback mass
\citep{2012ApJ...749...91F}. The direction of the natal kick is
assumed to be isotropic.

Our binary star model is based on the BSE model
\citep{2002MNRAS.329..897H}. We overview the parameters of our choice,
and the different points between the BSE and our models. The BSE model
includes wind accretion, tidal evolution, stable mass transfer, common
envelope evolution, magnetic braking, and orbital decay through GW
radiation. The wind accretion in our model is limited by the Eddington
limit expressed by \cite{1967Natur.215..464C}, while the wind
accretion in the BSE model is not. We adopt this limitation, since
wind accretion may exceed the Eddington limit when a star has a
massive companion. The BSE and our models have the same prescription
for tidal evolution of stars with convective envelopes. On the other
hand, we adopt two different prescriptions for tidal evolution of
stars with radiative envelopes, one of which is the same as the BSE
model (described in detail later). The BSE and our models have
different criteria for whether a star has radiative or convective
envelope. In the BSE model, a massive star has radiative (convective)
envelope at its core (shell) He burning phases, respectively. On the
other hand, in our model, a massive star has radiative or convective
envelope when its effective temperature ($\teff$) is above or below
$3.65$, respectively. The same criteria are applied also for the
stability of mass transfer. We need the different criteria, since the
M- and L-model stars can have radiative envelopes in their shell He
burning phases, and convective envelopes in their core He burning
phases unlike the BSE model. In a stable mass transfer phase, the
maximum fraction of transferred mass is $0.5$ in our model, while it
is unity in the BSE model. Such non-conservative mass transfer is also
adopted by the latest BPS calculations
\citep[e.g.][]{2020A&A...636A.104B, 2020MNRAS.498.3946K}. Our model as
well as the BSE model apply the $\alpha$ formalism for the common
envelope evolution \citep{1984ApJ...277..355W}. We set $\alpha=1$, and
adopt the formulae of \cite{2014A&A...563A..83C} for $\lambda$. We
assume that when a star in the Hertzsprung gap phase fills its Roche
lobe and its mass transfer is unstable, the star merges with its
companion star. This is because a star in its Hertzsprung gap phase
does not have steep density gradient between the He core and H
envelope \citep{2004ApJ...601.1058I}. This treatment is similar to
other binary population synthesis models \citep{2012ApJ...759...52D,
  2018MNRAS.474.2959G}. Note that such a star experiences common
envelope evolution in the BSE model.

We briefly describe the prescription for tidal evolution of stars with
radiative envelope, which is similar to \cite{2020MNRAS.498.3946K}. In
this case, the tidal evolution is subject to the radiative damping of
the dynamical tide \citep{1975A&A....41..329Z}. The tidal coupling
parameter $k/T$ is given by
\begin{align}
  \frac{k}{T} &= \left( \frac{Gm}{R^3} \right)^{1/2} (1+q)^{5/6}E
  \left( \frac{R}{a} \right)^{5/2} \\
  &= 2.9 \times 10^{-2} \; {\rm yr^{-1}} \; \left( \frac{m}{\msun}
  \right)^{1/2} \left( \frac{R}{\rsun} \right) \nonumber \\
  &\times \left( \frac{a}{{\rm au}} \right)^{-5/2} (1+q)^{5/6}
  E \label{eq:TidalSpinUp}
\end{align}
\citep[eq. (42) in ][see also \citealt{1977A&A....57..383Z,
    1981A&A....99..126H}]{2002MNRAS.329..897H}, where $R$ is the
stellar radius and $a$ is the binary semi-major axis. In the BSE
model, the variable $E$ depends on the stellar mass as
\begin{align}
  E = 1.1 \times 10^{-6} \left( \frac{m}{10 \msun}
  \right)^{2.84}. \label{eq:OriginalTide}
\end{align}
Recently, \citet[see also
  \citealt{2018A&A...616A..28Q}]{2010ApJ...725..940Y} proposed an
alternative fitting formula for the $E$ based on the ratio of the
convective core radius ($R_{\rm conv}$) to the stellar radius, such
that
\begin{align}
  E = \left\{
  \begin{array}{ll}
    10^{-0.42} \left( R_{\rm conv} / R \right)^{7.5} & \mbox{for H-rich stars} \\
    10^{-0.93} \left( R_{\rm conv} / R \right)^{6.7} & \mbox{for He-rich stars}
  \end{array}
  \right.. \label{eq:NewTide}
\end{align}
We set the convective core radius as
\begin{align}
  R_{\rm conv} \sim \left\{
  \begin{array}{ll}
    0.9 \rsun (Z/\zsun)^{0.05} (m/10\msun)^{0.8} & \mbox{for H-rich stars} \\
    0.5 \rsun & \mbox{for He-rich stars}
  \end{array}
  \right.. \label{eq:NewTideRconv}
\end{align}
We obtain the convective core radius for H-rich stars from our L-model
star's data. We adopt the same convective core radius for He-rich
stars as \cite{2020MNRAS.498.3946K}, although this value is fitted to
extremely metal-poor stars. Hereafter, the prescriptions for tidal
evolution expressed as Eqs. (\ref{eq:OriginalTide}) and
(\ref{eq:NewTide}) are called ``original-tide'' and ``new-tide''
models, respectively.

\subsection{Parameter sets}
\label{sec:ParameterSets}

We investigate eight parameter sets with different IMFs, binary star
initial conditions, and single and binary star models. We summarize
the parameter sets in Table \ref{tab:ParameterSets}.

We regard as the fiducial one the parameter set in which its IMF,
binary star initial condition, single star model, and binary star
model are the transitional IMF, the Sana's model, the M and SSE-Pop I
model with the standard PI model, and new-tide model, respectively. We
present the results of the SSE-std set only in appendices
\ref{sec:ComparisonWithSse} and \ref{sec:EffectsOfBinning}.

\begin{deluxetable*}{l|llll}
  \tablecaption{Summary of parameter sets. \label{tab:ParameterSets}}
  \tablehead{Parameter set & IMF & Binary IC & Single star & Binary star}
\startdata
Fiducial      & Transitional    & Sana              & M and SSE-Pop I with standard PI          & New tide \\
Top-light     & {\it Top-light} & Sana              & M and SSE-Pop I with standard PI          & New tide \\
Stepwise      & {\it Stepwise}  & Sana              & M and SSE-Pop I with standard PI          & New tide \\
Wide-binary   & Transitional    & {\it Wide binary} & M and SSE-Pop I with standard PI          & New tide \\
Original-tide & Transitional    & Sana              & M and SSE-Pop I with standard PI          & {\it Original tide} \\
L-std         & Transitional    & Sana              & {\it L and SSE-Pop I with standard PI}    & New tide \\
L-$\tsig$     & Transitional    & Sana              & {\it L and SSE-Pop I with $\tsig$ PI}     & New tide \\
SSE-std       & Transitional    & Sana              & {\it M and SSE-Pop I/II with standard PI} & New tide \\
\enddata
\tablecomments{We describe IMF, initial condition (IC) of binary
  stars, and single and binary star model for each model.  For the
  parameter sets other than the fiducial one, differences from the
  fiducial are accentuated by {\it italic}. }
\end{deluxetable*}

For parameter sets with the transitional IMF (e.g. the fiducial,
wide-binary, original-tide, L-std, L-$\tsig$, and SSE-std parameter
sets), we prepare $12$ groups of $10^6$ binary stars, and perform BPS
calculations for these groups. The $12$ groups include $10$ different
metallicities: $Z/\zsun = 10^{-8}$, $10^{-6}$, $10^{-4}$, $10^{-2}$,
$0.25$, $0.05$, $0.1$, $0.25$, $0.5$, and $1$. The two most metal-poor
groups follow the logarithmically flat IMF, and the six most
metal-rich groups follow the Kroupa's IMF. The two groups with
$Z/\zsun = 10^{-4}$ and $10^{-2}$ have a mixture of the Kroupa's and
logarithmically flat IMFs. For the top-light and stepwise parameter
sets, we prepare $10$ groups of $10^6$ binary stars with the $10$
different metallicities. In the top-light parameter set, all the
groups follow the Kroupa's IMF. For the stepwise parameter set, the
two most metal-poor groups follow the logarithmically flat IMF, and
the other groups follow the Kroupa's IMF.

We generate $10^6$ binary stars without any restriction on the upper
and lower mass limits in the case of the logarithmically flat IMF. The
total mass of these binary stars is $8.2 \times 10^7$ $\msun$. Since
we assume the binary number fraction to $0.5$ as described above, the
$10^6$ binary star formation should be accompanied by $10^6$ single
star formation. The total mass of the $10^6$ single stars and the
$10^6$ binary stars is $1.3 \times 10^8$ $\msun$. On the other hand,
we generate $10^6$ binary stars whose primary and secondary masses are
more than $10$ $\msun$ for the Kroupa's IMF. This is because the
number fraction of stars leaving behind BHs is small for the Kroupa's
IMF. The total mass of these binary stars is $4.2 \times 10^7$
$\msun$. These binary stars should be accompanied by a large number of
single stars and binary stars whose primary and secondary masses are
less than $10$ $\msun$. The total mass of the single and binary stars
(including the $10^6$ binary stars) is $5.2 \times 10^8$ $\msun$.

\subsection{BH merger rate density}
\label{sec:DifferentialBhMergerRateDensity}

We use the results of BPS calculations, and obtain the BH merger rate
density ($\rate$) and its derivatives by binary BH properties
($\prop$), such as primary mass, mass ratio, and their combination in
the source frame. First, we express the differential BH merger rate
density at a given look-back time as
\begin{align}
  \frac{d\rate}{d\prop} = \int dZ \int^{\thb-\tlb}_{0} d\td
  \frac{d\phi(z(\td+\tlb))}{dZ} \frac{d^{2}\neff(Z)}{d\td
    d\prop}, \label{eq:dmrd}
\end{align}
where $\thb$ is the Hubble time, $\tlb$ is the look-back time, and
$\td$ is the delay time of a BH merger from star formation. We can
regard $\neff(Z)$ as the total number of BH mergers per stellar mass
formed at a given metallicity $Z$. We call $\neff(Z)$ ``production
efficiency of BH mergers''.\footnote{The production efficiency of BH
  mergers $\neff(Z)$ is similar to, but slightly different from the
  merger efficiency defined by \cite{2018MNRAS.480.2011G} and the
  formation efficiency defined by \cite{2018A&A...619A..77K}. This is
  because $\neff(Z)$ includes BH mergers whose delay time is infinity,
  or more than the Hubble time by definition. In section
  \ref{sec:FiducialParameterSet}, we define the production efficiency
  of BH mergers over the Hubble time $\nint(Z)$, which is identical to
  the merger or formation efficiency.}

We discretize the integral of $Z$ in Eq. (\ref{eq:dmrd}), and obtain
the following equation:
\begin{align}
  \frac{d\rate}{d\prop} \approx \sum_{i=1}^{N_{\rm Z}}
  \int^{\thb-\tlb}_{0} d\td \Delta \phi_i(z(\td+\tlb))
  \frac{d^{2}\neff (Z_i)}{d\td d\prop}. \label{eq:dmrdz}
\end{align}
$N_{\rm Z}$ is the number of different metallicities (i.e. $10$), and
$Z_i/\zsun = 10^{-8}, 10^{-6}, \cdots, 0.5, 1$ for $i=1, 2, \cdots, 9,
10$, respectively. $\Delta \phi_i (z)$ is the SFR density at a given
redshift $z$ in a metallicity range around metallicity $Z_i$. The
lower and upper boundaries of the metallicity range are the geometric
means with its lower and upper neighbor metallicities:
$(Z_{i-1}Z_i)^{1/2}$ and $(Z_iZ_{i+1})^{1/2}$, respectively. We define
$Z_0=0$ and $Z_{11}=\infty$. Thus, we can calculate $\Delta \phi_i
(z)$ as
\begin{align}
  \Delta \phi_i (z) &= \int_{(Z_{i-1}Z_i)^{1/2}}^{(Z_iZ_{i+1})^{1/2}}
  dZ \frac{d\phi(z)}{dZ},
\end{align}
where $d\phi(z)/dZ$ can be expressed as Eq. (\ref{eq:SfrDiffZ}). Note
that we use the $Z=Z_\odot$ result for this metallicity range,
although $(Z_iZ_{i+1})^{1/2}$ is infinite for $i=10$.

We can also discretize the integral in Eq. (\ref{eq:dmrdz}) as
follows. The derivative in Eq. (\ref{eq:dmrdz}) can be divided into
two parts: binary stars subject to the Kroupa's and logarithmically
flat IMFs, such that
\begin{align}
  \frac{d^{2}\neff (Z_i)}{d\td d\prop} = \sum_{j=1}^{N_{\rm IMF}}
  f_{i,j} \frac{d^{2}\neff_{i,j}}{d\td d\prop}, \label{eq:PartOfDiff}
\end{align}
where $N_{\rm IMF}$ is the number of IMFs (i.e. $N_{\rm IMF}=1$ or
$2$), $f_{i,j}$ is the mass fraction of stars formed in each IMF, and
$\neff_{i,j}$ is production efficiency of BH mergers for each $Z_i$
and each IMF. Note that $f_{i,1}+f_{i,2}=1$. Eq. (\ref{eq:PartOfDiff})
can be written as
\begin{align}
  \frac{d^{2}\neff_{i,j}}{d\td d\prop} = \frac{1}{M_{{\rm
        bps},i,j}} \sum_{k=1}^{N_{{\rm bps},i,j}} \delta(\td -
  \td^{i,j,k}) \delta(\prop - \prop^{i,j,k}), \label{eq:DiscretOfDiff}
\end{align}
where $N_{{\rm bps},i,j}$ is the number of simulated binary stars for
each $Z_i$ and each IMF (i.e. $N_{{\rm bps},i,j}=10^6$), $M_{{\rm
    bps},i,j}$ is the total stellar mass required for forming the
$N_{{\rm bps},i,j}$ binary stars (i.e. $M_{{\rm bps},i,1} = 5.2 \times
10^8$ $\msun$ and $M_{{\rm bps},i,2} = 1.3 \times 10^8$ $\msun$ as
described in section \ref{sec:ParameterSets}), and $\td^{i,j,k}$ and
$\prop^{i,j,k}$ are the delay time and binary BH properties of each
binary star. Using Eqs. (\ref{eq:PartOfDiff}) and
(\ref{eq:DiscretOfDiff}), we can discretize Eq. (\ref{eq:dmrdz}) as
\begin{align}
  \frac{d\rate}{d\prop} \approx \sum_{i=1}^{N_{\rm Z}}
  \sum_{j=1}^{N_{\rm IMF}} \sum_{k=1}^{N_{{\rm bps},i,j}}
  \frac{f_{i,j} \Delta \phi_i(z(\td^{i,j,k}+\tlb))}{M_{{\rm bps},i,j}}
  \delta(\prop - \prop^{i,j,k}).
\end{align}
The BH merger rate density can also be obtained as
\begin{align}
  \rate \approx \sum_{i=1}^{N_{\rm Z}} \sum_{j=1}^{N_{\rm IMF}}
  \sum_{k=1}^{N_{{\rm bps},i,j}} \frac{f_{i,j} \Delta
    \phi_i(z(\td^{i,j,k}+\tlb))}{M_{{\rm bps},i,j}}.
\end{align}

\section{Results}
\label{sec:Results}

We present the results for the fiducial parameter set in section
\ref{sec:FiducialParameterSet}. We see how the results change in the
cases with other parameter sets in section
\ref{sec:OtherParameterSets}.

\subsection{Fiducial parameter set}
\label{sec:FiducialParameterSet}

\begin{figure}[ht!]
  \plotone{\fdir/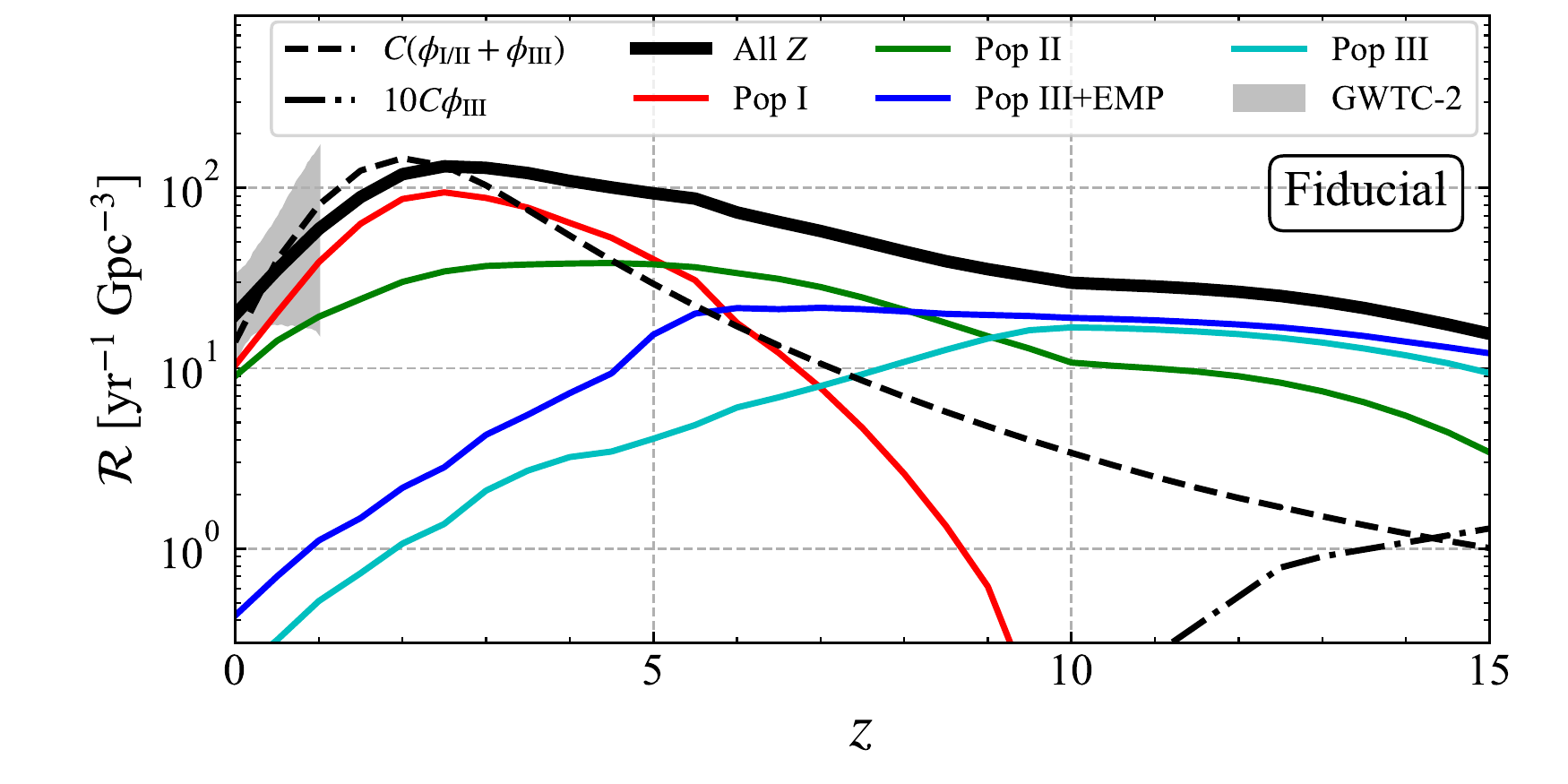}
  \caption{Redshift evolution of the BH merger rate density in the
    case with the fiducial parameter set. Black, red, green, blue, and
    cyan curves indicate the BH merger rate densities for all $Z$, Pop
    I, Pop II, Pop III+EMP, and Pop III, respectively. The gray shaded
    region shows the 90\% credible interval inferred by the LIGO-Virgo
    Gravitational-Wave Transient Catalog 2 (GWTC-2). The dashed curve
    indicates the redshift evolution of the sum of the Pop I/II and
    Pop III SFRs ($\phi_{\rm I/II}$ and $\phi_{\rm III}$,
    respectively) scaled by an arbitrarily value $C$. The
    dashed-dotted curve indicates the redshift evolution of the Pop
    III SFR scaled by $10C$.}
  \label{fig:fdcl_mergerRateTotal}
\end{figure}

Figure \ref{fig:fdcl_mergerRateTotal} shows the redshift evolution of
the BH merger rate density for the fiducial set. We find ${\cal R}
\sim 20$ yr$^{-1}$ Gpc$^{-3}$ at $z=0$, comparable to the observed
rate, $19.3^{+15.1}_{-9.0}$ yr$^{-1}$ Gpc$^{-3}$ according to the
LIGO-Virgo Gravitational-Wave Transient Catalog 2 (GWTC-2)
\citep{2021PhRvX..11b1053A}\footnote{We refer to the BH merger rate
  density, and BH and spin mass distribution based on GWTC-2
  \citep{2021ApJ...913L...7A}, not the latest one on GWTC-3
  \citep{2021arXiv211103634T}.}. Note that our prediction is also
consistent with the GWTC-2.1 result \citep{2021arXiv210801045T}. The
predicted merger rate density up to $z=1$ also falls within the $90$
\% credible interval inferred by GWTC-2.

The BH merger rate density monotonically increases with increasing
redshift up to $z \simeq 2.5$, and then decreases monotonically for
higher redshift. The BH merger rate density is dominated Pop I/II
stars at redshifts of $z \lesssim 7$.  Pop III and EMP stars
contribute only by $\sim 1$ ($10$) \% to the BH merger rate density at
the redshift of $z = 0$ ($z=5$, respectively). Their contribution
becomes dominant at redshifts higher than $z \sim 7$.

We find that the redshift at which the BH merger rate reaches the
maximum ($z \sim 2.5$) is slightly higher than the redshift of the
maximum SFR ($z \sim 2$). In naive expectations, the BH merger rate
should achieve the maximum at some epoch after the peak SFR as binary
BHs merge after their progenitor formation. However, this is not the
case for the following reason. At $z < 3$, the SFR for $Z \sim 0.25$
$\zsun$ is lower than that for $Z \sim 0.5$ ($1$) $\zsun$ by a factor
of about $3$ ($10$). On the other hand, the production efficiency of
BH mergers over the Hubble time ($\nint(Z) = \int_0^{\thb} d\td
d\neff(Z)/d\td$)\footnote{As described in the previous footnote, the
  production efficiency of BH mergers over the Hubble time $\nint(Z)$
  is the same as the merger efficiency called by
  \cite{2018MNRAS.480.2011G} and the formation efficiency called by
  \cite{2018A&A...619A..77K}.} for $Z \sim 0.25$ $\zsun$ is higher
than those for $Z \sim 0.5$ and $1$ $\zsun$ by a factor of about $30$
and $1000$, respectively, such that $\nint(Z) = 9.1 \times 10^{-6}$
$\msun^{-1}$ for $Z=0.25$ $\zsun$, $3.0 \times 10^{-7}$ $\msun^{-1}$
for $Z=0.5$ $\zsun$, and $7.6 \times 10^{-9}$ $\msun^{-1}$ for $Z=1$
$\zsun$. Such sharp dependence on metallicity has also been pointed
out by \cite{2018MNRAS.480.2011G} and
\cite{2018A&A...619A..77K}. Merging binary BHs are thus mainly yielded
by stars of $Z \simeq 0.25$ $\zsun$ at $z < 3$. Moreover, about half
of them have much shorter inspiral times than the Hubble time
($\lesssim 100$ Myr). The redshift evolution of the BH merger rate
density follows the SFR of $Z \simeq 0.25$ $\zsun$ stars, which
achieves the maximum at $z \simeq 2.5$.

\begin{figure}[ht!]
  \plotone{\fdir/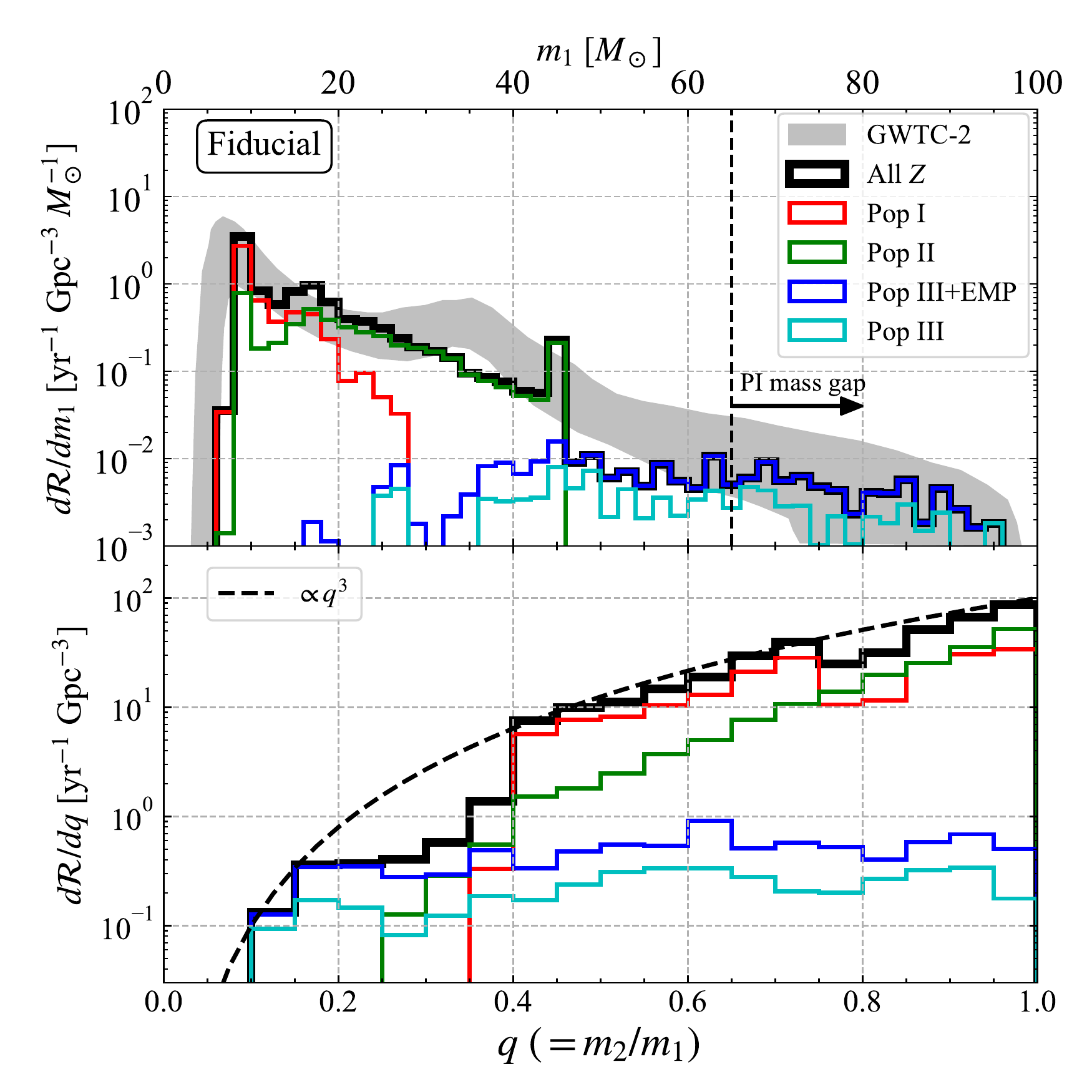}
  \caption{BH merger rate density differentiated by primary BH mass
    (top) and mass ratio (bottom) at the redshift of $z=0$ in the case
    with the fiducial set. The color codes are the same as Figure
    \ref{fig:fdcl_mergerRateTotal}. The PI mass gap region is on the
    right-hand side of the dashed line in the top panel. The dashed
    curve in the bottom panel indicates $d\rate/dq \propto q^{3}$.}
  \label{fig:fdcl_mergerRateMassMetal}
\end{figure}

We compare our obtaining BH masses with BH masses inferred by GWTC-2
\citep{2021ApJ...913L...7A}. Figure \ref{fig:fdcl_mergerRateMassMetal}
shows the distributions of the primary BH masses (top) and mass ratios
(bottom) at $z=0$, which are defined as the BH merger rate density
differentiated by the primary BH mass and mass ratio, respectively.
The primary BH mass distribution is roughly consistent with the
observations in GWTC-2 in three respects. First, our primary BH mass
distribution has the global maximum at $m_1 \sim 10$ $\msun$. Second,
the mass distribution suddenly decreases at $m_1 \sim 45$ $\msun$ due
to the PPI effect. It should be noted that merging binary BHs with
$m_1 \sim 45$ $\msun$ are overproduced while those with $m_1 \sim 35$
and $50$ $\msun$ are underproduced, relative to GWTC-2, because of the
standard PI modeling in which PPI produces BHs with a fixed mass ($45$
$\msun$). If we had adopted a PPI model whose remnant mass has a broad
distribution \citep[e.g.][]{2016MNRAS.457..351Y, 2019ApJ...887...72L,
  2020ApJ...905L..21U}, this discrepancy would have
disappeared. Third, we find BHs within the PI mass gap, $m_1 \sim
65$--$90$ $\msun$. As shown in \cite{2021MNRAS.505.2170T}, Pop III
stars of $65$--$90$ $\msun$ keep their small radii ($\lesssim 40$
$\rsun$), and do not lose their H envelopes through binary
interactions. They have He cores with $\lesssim 45$ $\msun$, and thus
leave behind $65$--$90$ $\msun $ BHs without either PPI or PISN
effect. As seen in comparison between the blue and cyan curves in
Figure \ref{fig:fdcl_mergerRateMassMetal}, not only Pop III binary
stars but also EMP binary stars contribute to PI mass gap mergers.
The results in Figure \ref{fig:fdcl_mergerRateMassMetal} also
eliminate our concerns described in section \ref{sec:Introduction}. In
other words, Pop I/II stars can produce binary BHs in the mass range
of $\lesssim 50$ $\msun$, and do not overproduce the PI mass gap
events.

We can see that only Pop III and EMP stars can produce merging binary
BHs within the PI mass gap. In the mass range $65$--$90$ $\msun$,
their radii remain small, and thus they can keep their H envelope
until the collapse to BHs. They thus leave behind the PI mass gap
BHs. On the other hand, Pop I/II stars in the mass range $65$--$90$
$\msun$ expand greatly, and lose their H envelope through binary
interactions. They leave behind at most $\sim 45$ $\msun$ BHs. The
radius evolution can be seen in the Appendix
\ref{sec:SingleStarModels}. In advance, we should remark that the
formation of merging binary BHs within the PI mass gap strongly
depends on the choice of the single star model for Pop III and EMP
stars. If we choose the L model, we do not have such binary BHs as
seen in section \ref{sec:OtherParameterSets}.

The bottom panel of Figure \ref{fig:fdcl_mergerRateMassMetal} shows
the mass ratio distribution of merging binary BHs. The distribution
roughly follows the power law $\propto q^{3}$ (dashed
curve). According to \cite{2021ApJ...913L...7A}, the power-law index
of the GWTC-2 events is less than or equal to $3$. Our mass ratio
distribution is thus marginally consistent with it.

We find that Pop I/II binary stars preferentially result in equal mass
ratio merger events of $q \sim 1$.  Because of the large radii in the
post-MS phase, they tend to experience binary interactions, such as
common envelope and stable mass transfer.  After the binary
interactions, which work to equalize the binary component masses, they
have similar BH masses.  On the other hand, Pop III and EMP stars
hardly experience such binary interactions, and keep their initial
mass ratios and thus the initial flat $q$ distribution (see section
\ref{sec:InitialConditions}).  In fact, their mass ratios follow
nearly flat distribution as seen in the lower panel in Figure
\ref{fig:fdcl_mergerRateMassMetal} (blue).

\begin{figure}[ht!]
  \plotone{\fdir/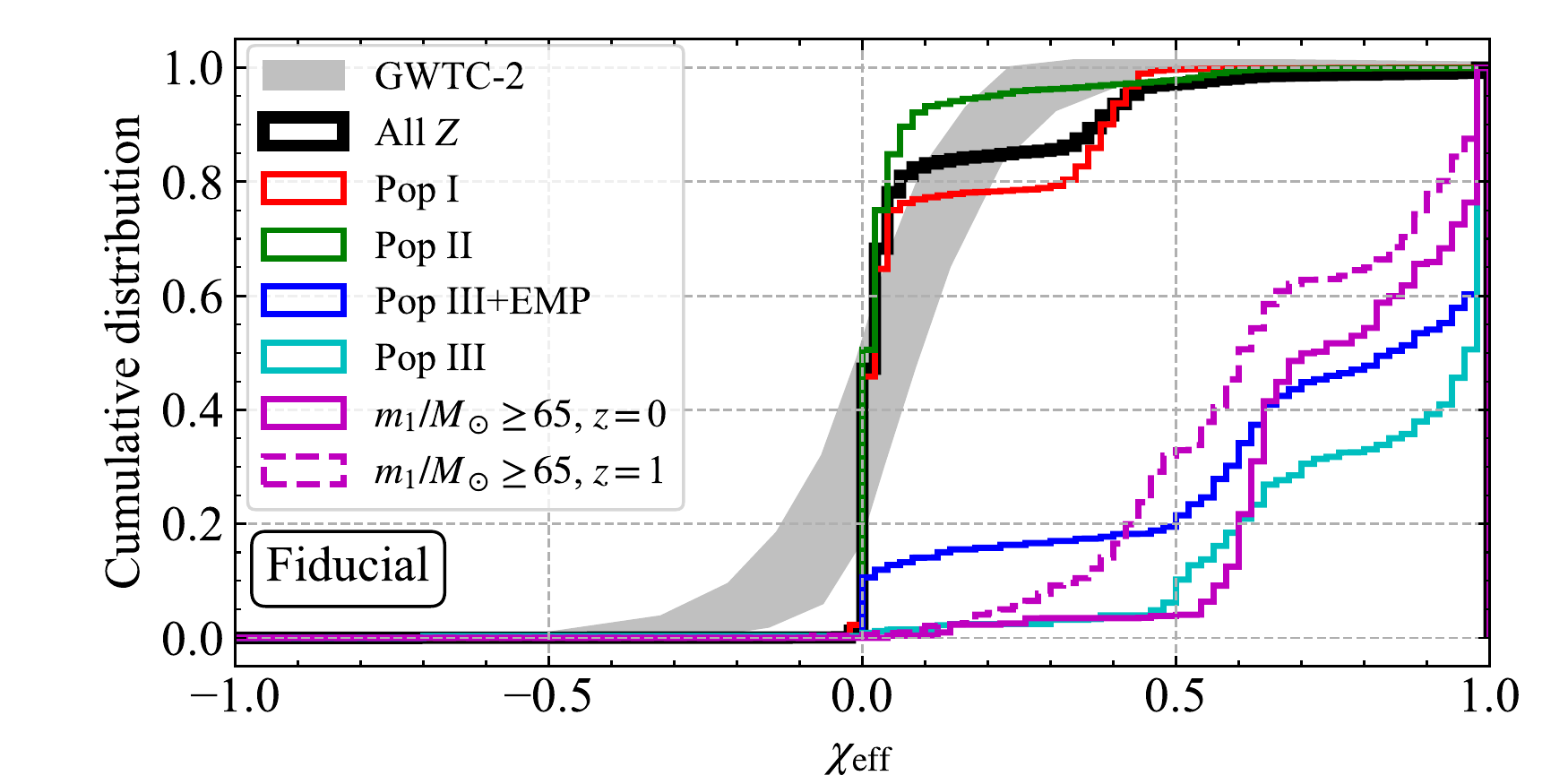}
  \caption{Cumulative distribution of effective spin at the redshift
    of $z=0$ for the fiducial set. The color codes are the same as
    Figure \ref{fig:fdcl_mergerRateTotal}. The magenta solid and
    dashed curves indicate the cumulative distribution for $m_1/\msun
    \ge 65$ at $z=0$ and $1$, respectively.}
  \label{fig:fdcl_mergerRateXeffMetal}
\end{figure}

Figure \ref{fig:fdcl_mergerRateXeffMetal} shows the cumulative
distribution of effective spins for the fiducial set of the
parameters.  The effective spin of a binary BH system is defined as
\begin{align}
  \xeff = \frac{m_1 \chi_1 \cos \theta_1 + m_2 \chi_2 \cos
    \theta_2}{m_1+m_2},
\end{align}
where $\chi_1$ and $\chi_2$ are the dimensionless spin of the primary
and secondary BHs, respectively, and $\theta_1$ and $\theta_2$ are the
inclination angle between the primary and secondary spin vectors and
the orbital angular momentum vector of the binary BH,
respectively. All of merging binary BHs have zero or positive
$\xeff$. This means that they have either no spin or spins aligned to
their orbital angular momentum vectors, since they have $\theta_1 \sim
0$ and $\theta_2 \sim 0$. Such spinning BHs acquire their spin angular
momenta in the progenitor phase through tidal interactions with the
companion stars.

About $10$ \% of merging BHs have $\xeff \sim 0.4$, which are formed
from Pop I binary stars in the following pathway.  After the primary
star in a binary system collapses to a BH with little spin, the
secondary star becomes a giant, and fills its Roche lobe. The binary
system then experiences the common envelope evolution, and shrinks its
separation down to $\lesssim 10$ $\rsun$. The secondary star loses its
H envelope and eventually becomes a naked He star. Because of the
small separation, it is efficiently spun up by the tidal interaction
(see the formulae in Eqs. (\ref{eq:TidalSpinUp}), (\ref{eq:NewTide}),
and (\ref{eq:NewTideRconv})), and collapses to a BH conserving its
spin angular momentum (see Eqs. (\ref{eq:CollapsingBhSpin}) and
(\ref{eq:BhSpin})). Since the resultant binary BHs have small
separation, they merge after a short delay time of $\sim 100$ Myr from
their progenitor star formation. This has been argued by
\cite{2016MNRAS.462..844K} and \cite{2017ApJ...842..111H}.

Pop II binary stars leave behind merging BHs with a wide range of
positive $\xeff \sim 0$--$1$.  These binary BHs especially dominate
the population of merging BHs with $\xeff \gtrsim 0.4$. The reason for
the wide range of $\xeff$ is because their progenitor stars are spun
up by tidal interactions, and lose their H envelope, i.e., large
reservoir of spin angular momenta, to some extent through stable mass
transfer. Note that their progenitors lose little mass through stellar
winds because of their low metallicity, and tend not to experience the
common envelope evolution since they become blue supergiant stars (see
appendix \ref{sec:SingleStarModels}). The spin magnitudes of these
binary BHs depend on whether their progenitors keep the H envelope.

Most of Pop III and EMP binary stars also leave behind merging BHs
with higher $\xeff$ than Pop II binary stars.  Pop III and EMP stars
with smaller radii, tend to keep more massive H envelope than more
metal-enriched stars (see appendix
\ref{sec:SingleStarModels}). Merging BHs resulting from Pop III and
EMP binary stars however, do not much affect the cumulative $\xeff$
distribution of all the merging binary BHs. This is because they
account for only $1$ \% of the total BH merger rate density at $z=0$.

There are, however, discrepancies between our cumulative
$\xeff$-distribution and observations in three respects. First, no
binary BH with negative $\xeff$ is found in our sample. Second, our
binary BHs are fewer than observed in the range $\xeff \sim 0$--$0.1$.
Third, there are more binaries BHs with $\xeff \sim 0.4$ in our sample
than what is observed.  We nevertheless expect that we can mitigate
these deviations once we account for observational uncertainties as in
\cite{2020A&A...635A..97B}.

Figure \ref{fig:fdcl_mergerRateXeffMetal} also shows the cumulative
distribution of binary BHs with $m_1/\msun \ge 65$, which are within
the PI mass gap, at $z=0$ and $1$. More than $90$ and $70$ \% of
binary BHs within the PI mass gap have $\xeff > 0.5$ at the redshifts
of $z=0$ and $1$, respectively.  The events at higher redshift have
smaller $\xeff$ since the progenitor binary stars have smaller
separations at higher redshift, and lose their H envelope more
promptly.

We have presented the cumulative distribution of binary BHs with PI
mass gap BHs both for $z=0$ and $1$, since some of the observed PI
mass gap BHs are located closer to $z=1$ rather than $z=0$. Here, we
regard the observed binary BHs with $m_1/\msun \ge 65$ as PI mass gap
events even if they are not conclusive. For example, GW190521 is
conclusive, since its $m_1$ is in the PI mass gap with the $90$ \%
credible intervals \citep{2020PhRvL.125j1102A,
  2020ApJ...900L..13A}. On the other hand, the followings are not:
GW190519\_153544, GW190602\_175927, GW190706\_222641, and
GW190929\_012149 \citep{2021PhRvX..11b1053A}, and GW190403\_051519 and
GW190426\_190642 \citep{2021arXiv210801045T}. The conclusive event,
GW190521, has small $\xeff$ ($0.03^{+0.32}_{-0.39}$), although the
$90$ \% credible interval is quite large. On the other hand, the
non-conclusive ones have a wide variety of $\xeff$. For example,
GW190403\_051519, has high $\xeff$ ($0.70^{+0.15}_{-0.27}$ with the
$90$ \% credible intervals). This suggests that our model may still be
consistent with the observation even considering the high probability
of PI mass gap BHs with high spins in our sample.

We do not try to compare our model with the observations in terms of
the so-called spin precession, i.e., the BH spin component
perpendicular to the binary orbital angular momentum vector. This is
because the estimates of spin precessions have larger uncertainties
than those of effective spins, even though meaningful constraints may
be imposed on spin precessions of a few events in GWTC-2 data
\citep{2021PhRvX..11b1053A}, such as GW190412
\citep{2020PhRvD.102d3015A}.

\begin{figure}[ht!]
  \plotone{\fdir/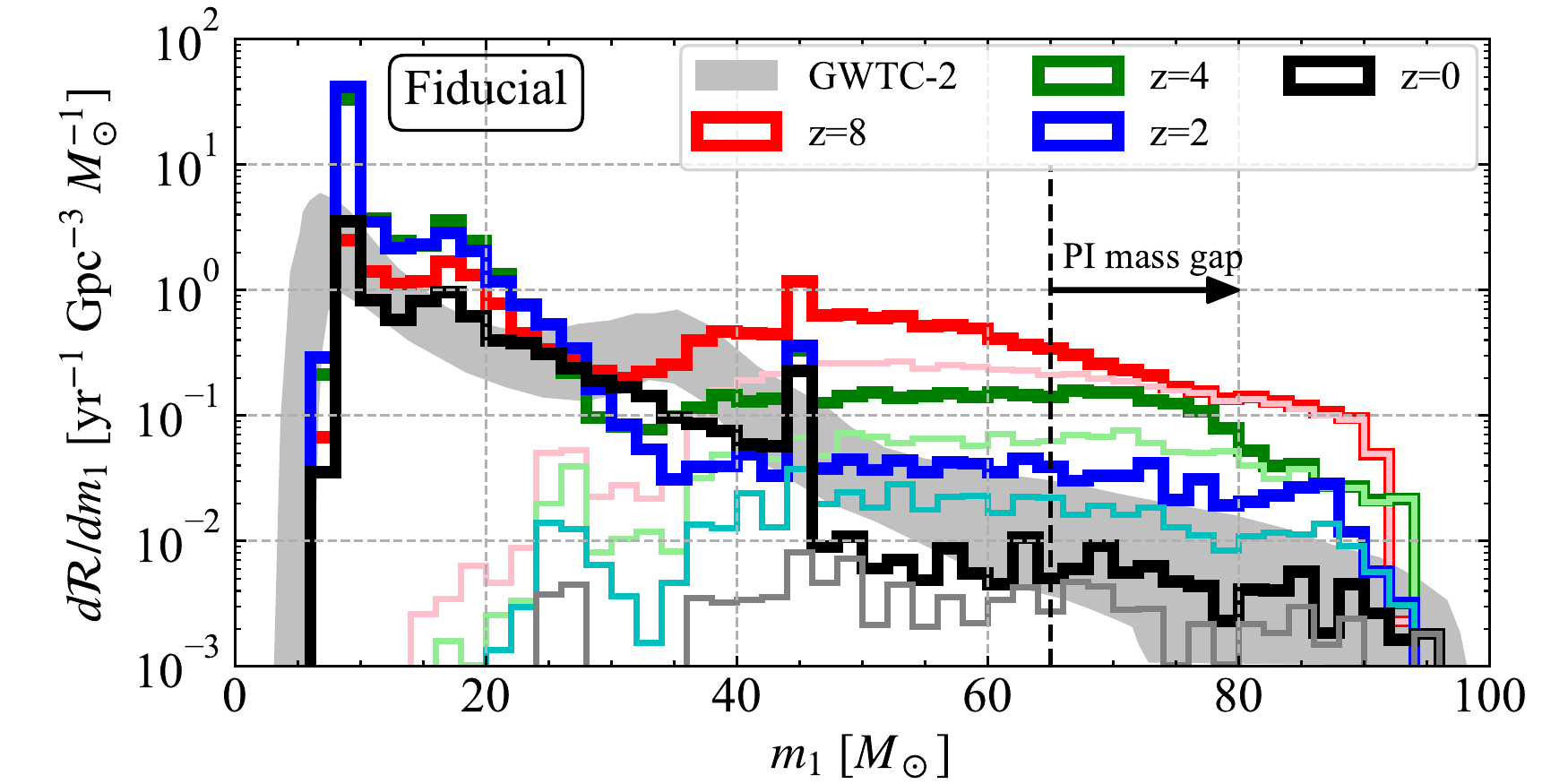}
  \caption{BH merger rate density as a function of primary BH mass at
    the redshifts of $0$ (black), $2$ (blue), $4$ (green), and $8$
    (red) for the fiducial set. The gray, cyan, light-green, and pink
    curves indicate Pop III contributions at the redshifts of $0$,
    $2$, $4$, and $8$, respectively. The shaded gray regions show the
    90\% credible interval inferred by the LIGO-Virgo
    Gravitational-Wave Transient Catalog 2 (GWTC-2), which is relevant
    for the local redshift, $z<2$. The PI mass gap region is on the
    right-hand side of the dashed line.}
  \label{fig:fdcl_mergerRateMassRedshift}
\end{figure}

Figure \ref{fig:fdcl_mergerRateMassRedshift} shows the redshift
evolution of the primary BH mass distribution for the fiducial set of
the parameters. The global maximum at $m_1 \sim 10$ $\msun$ increases
from $z=0$ to $z=2-4$ by an order of magnitude, and then again
decreases from $z=4$ to $z=8$ by an order of magnitude. This redshift
evolution reflects that of the Pop I BH merger rate. Note that merging
binary BHs around the global maximum results from the Pop I
progenitors. We find that the BH merger rate density at $m_1 \sim 10$
$\msun$ grows much more than at $m_1 \sim 15$ $\msun$ from $z=0$ to
$z=2$ and $4$, despite that Pop I binary stars can yield binary BHs
with both $m_1 \sim 10$ and $15$ $\msun$ at the redshift of $z=0$ (see
Figure \ref{fig:fdcl_mergerRateMassMetal}). This makes a spike in the
BH merger rate density at $m_1 \sim 10$ $\msun$ at $z=2$ and $4$.
This spike comes from different formation pathways between Pop I
binary BHs with $m_1 \sim 10$ and $15$ $\msun$. The former binary BHs
are formed through common envelope evolution, and typically have short
delay time of $\sim 100$ Myr from their progenitor star formation. On
the other hand, the latter binary BHs are formed through stable mass
transfer, and their typical merger time is $\sim 10$ Gyr. Thus, the
former merger rate density grow much more than the latter from $z=0$
to $z=2$ and $4$. In advance, we note that this feature can be seen in
all the sets (see Figure \ref{fig:mergerRateMassRedshift}).

Interestingly, the BH merger rate density in the mass range $m_1 \sim
20$--$30$ $\msun$ does not change over the time interval
$z=0$--$8$. This is because they are originated mainly from Pop II
binary stars (see Figure \ref{fig:fdcl_mergerRateMassMetal}), and
their merger rate density does not change over those redshifts (see
Figure \ref{fig:fdcl_mergerRateTotal}).

The merger rate density of binary BHs within the PI mass gap
($\sim65$--$90$ $\msun$) monotonically increases up to $z=8$.  They
are formed from the Pop III and EMP progenitors (see Figure
\ref{fig:fdcl_mergerRateMassMetal}), whose merger rate increases with
redshift up to $z \sim 8$ (see Figure
\ref{fig:fdcl_mergerRateTotal}). This is because the Pop III SFR
increases toward higher redshifts up to $z \sim 20$.

We can see from Figure \ref{fig:fdcl_mergerRateMassRedshift} that Pop
III contributions in the PI mass gap become larger with increasing
redshift. At $z=0$, EMP binary stars contribute to the PI mass gap
equally to Pop III binary stars. However, Pop III binary stars start
dominating the PI mass gap from above. At $z=2$, $4$, and $8$, Pop III
binary stars have great roles on forming binary BHs with $m_1 \gtrsim
90$, $80$, and $70$ $\msun$ in the mass gap, respectively. We can
expect that a mass gap event has its origin in a Pop III binary star
if it is detected at a redshift $z \gtrsim 8$.

\begin{figure}[ht!]
  \plotone{\fdir/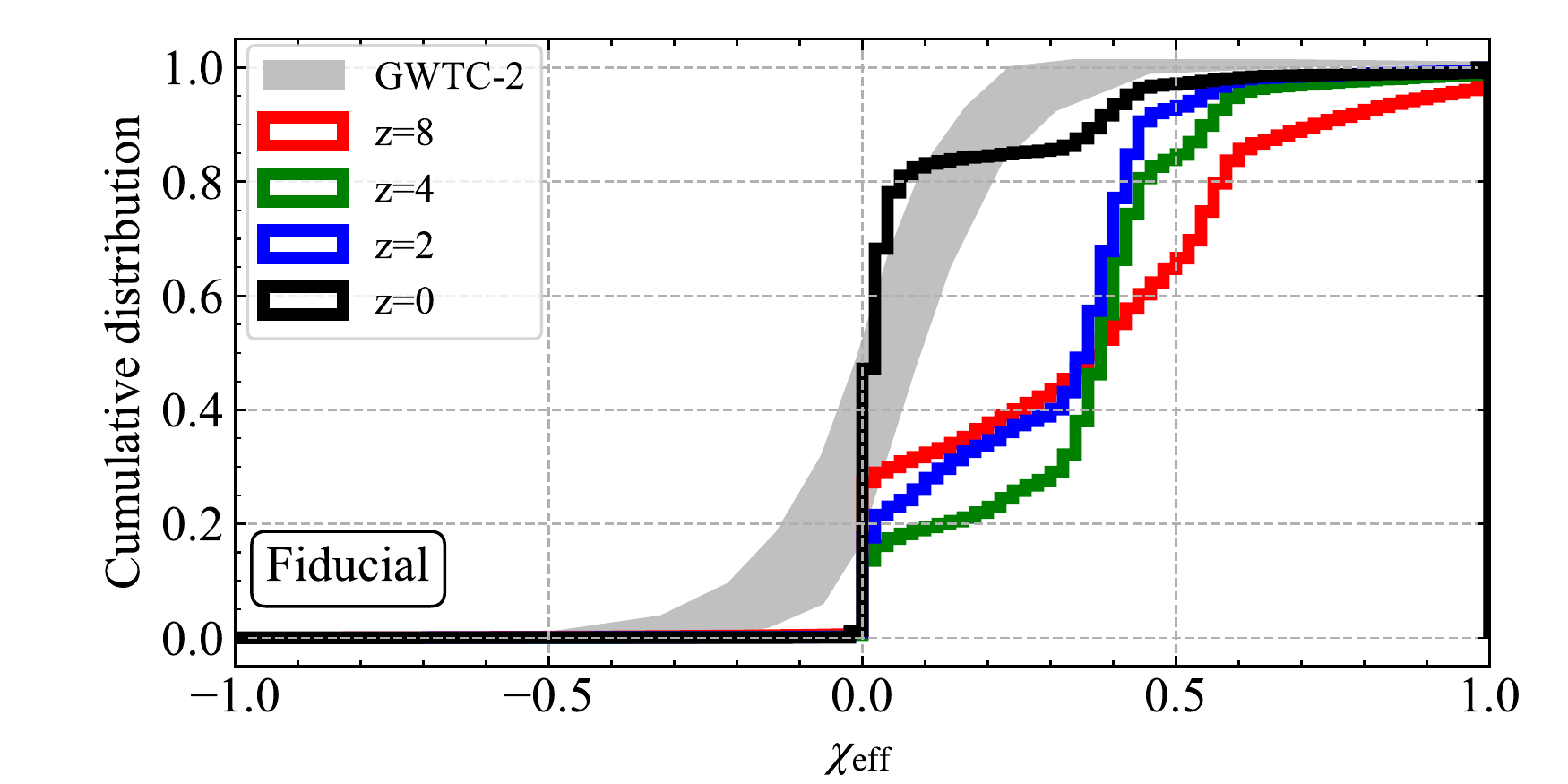}
  \caption{Cumulative distribution of effective spins for the fiducial
    set. The color codes are the same as Figure
    \ref{fig:fdcl_mergerRateMassRedshift}.}
  \label{fig:fdcl_mergerRateXeffRedshift}
\end{figure}

Figure \ref{fig:fdcl_mergerRateXeffRedshift} shows the redshift
evolution of the cumulative effective spin distribution for the
fiducial set. The contribution of binary BHs with $\xeff \sim 0.4$
increases from $z=0$ to $z=4$. As described above, they are originated
from Pop I binary stars, and have short delay time since their
progenitor formation. Thus, their merger events occur more frequently
at redshifts when star formation (especially of Pop I stars) is more
active (i.e. $z=2$ and $4$). At $z=8$, the fraction of binary BHs with
$\xeff \simeq 0.4$ is low, while the fraction with higher $\xeff$ is
high. This is because at this redshift binary BHs are originated
dominantly from the Pop II, Pop III and EMP progenitors, especially
Pop III and EMP progenitors.

\subsection{Other parameter sets}
\label{sec:OtherParameterSets}

\begin{figure*}[ht!]
  \plottwo{\fdir/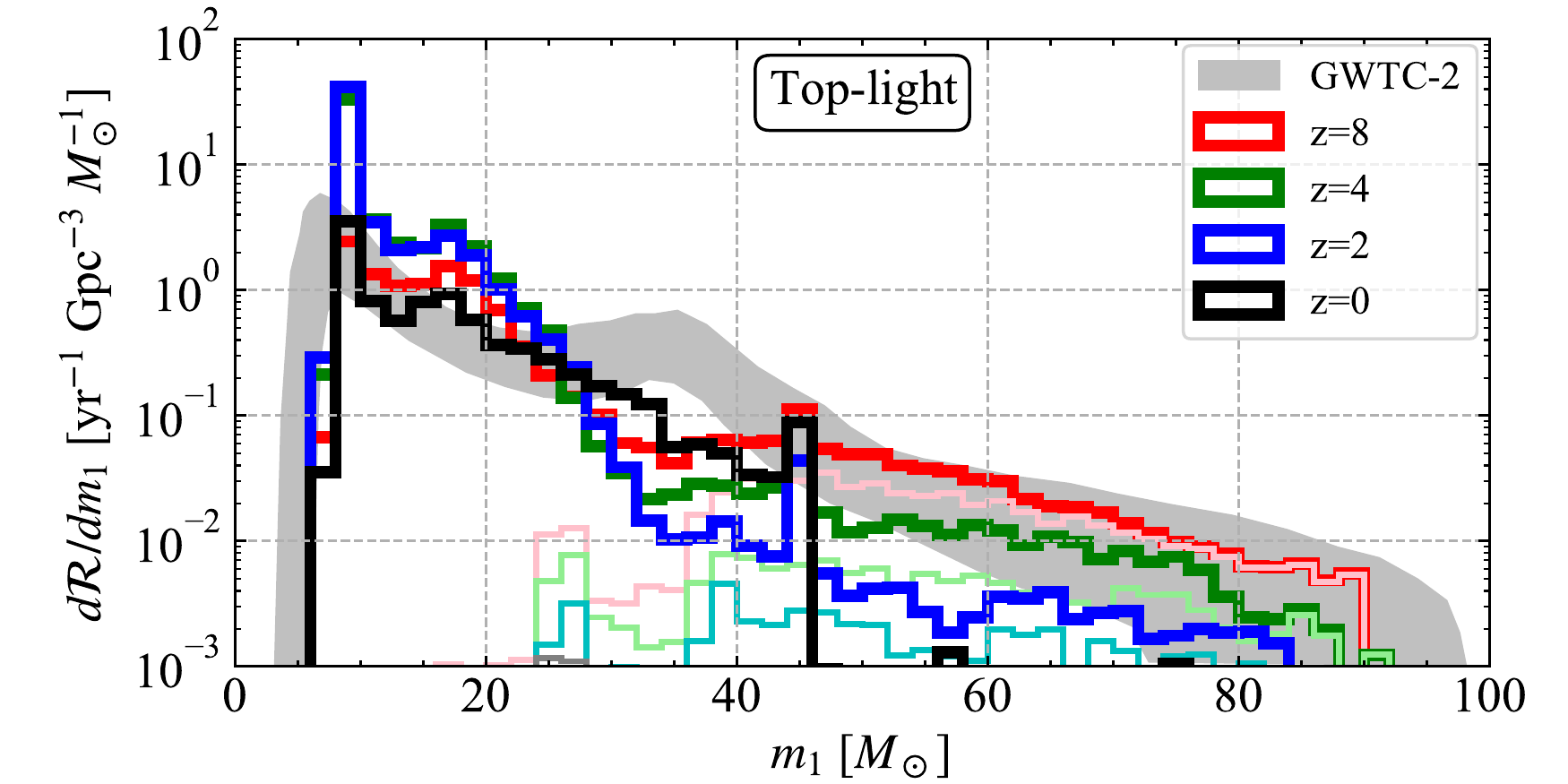}{\fdir/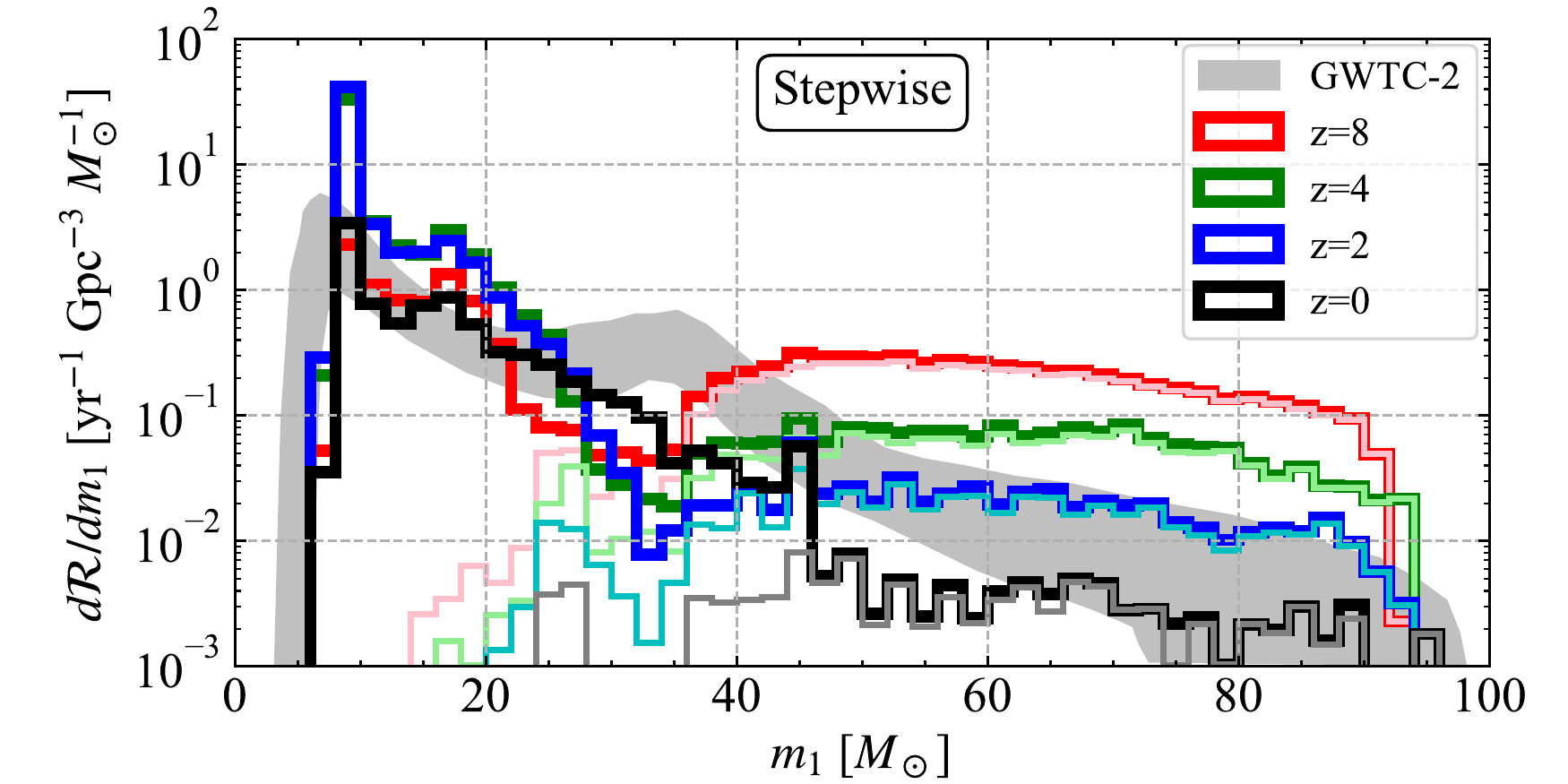}
  \plottwo{\fdir/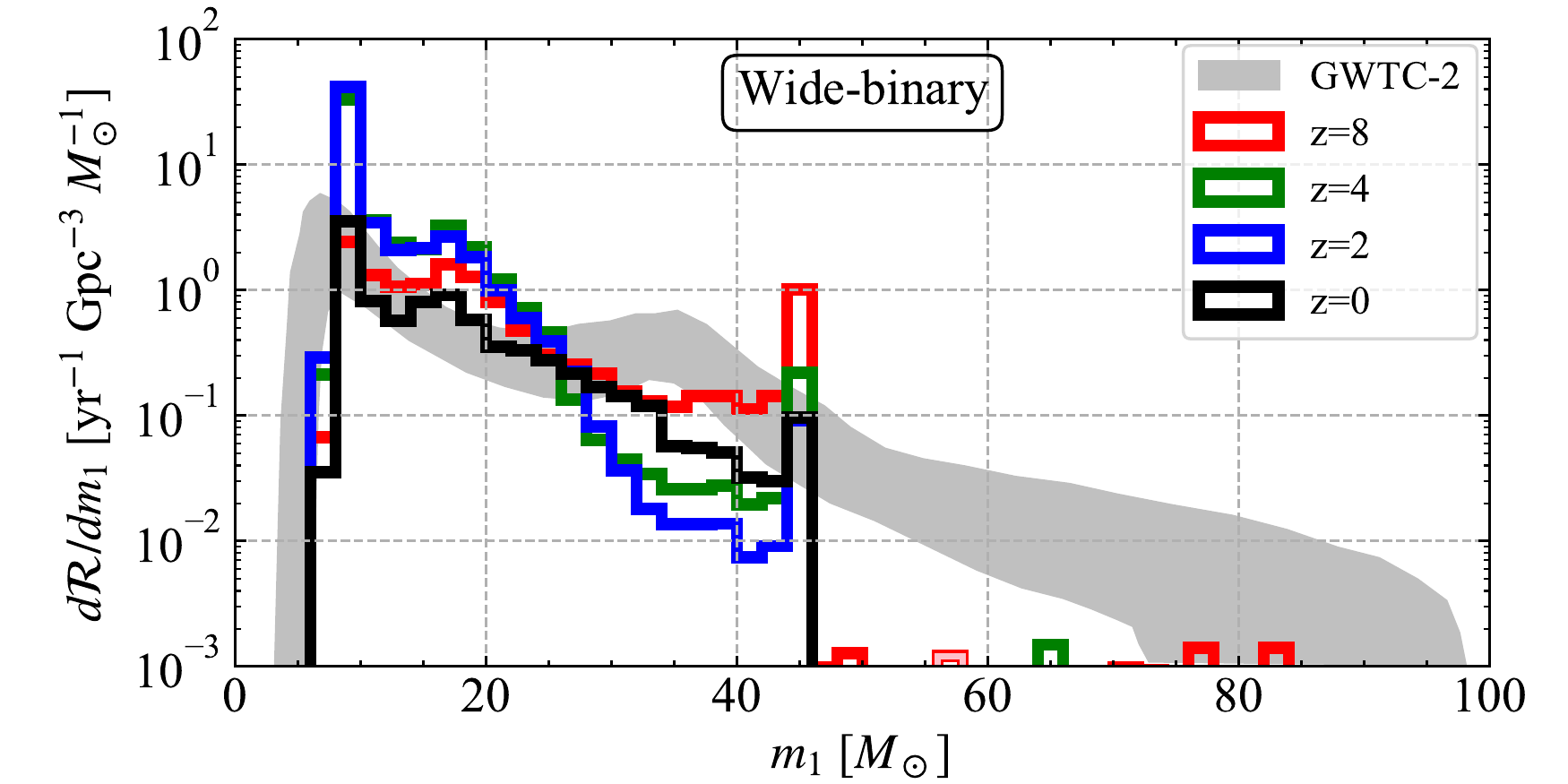}{\fdir/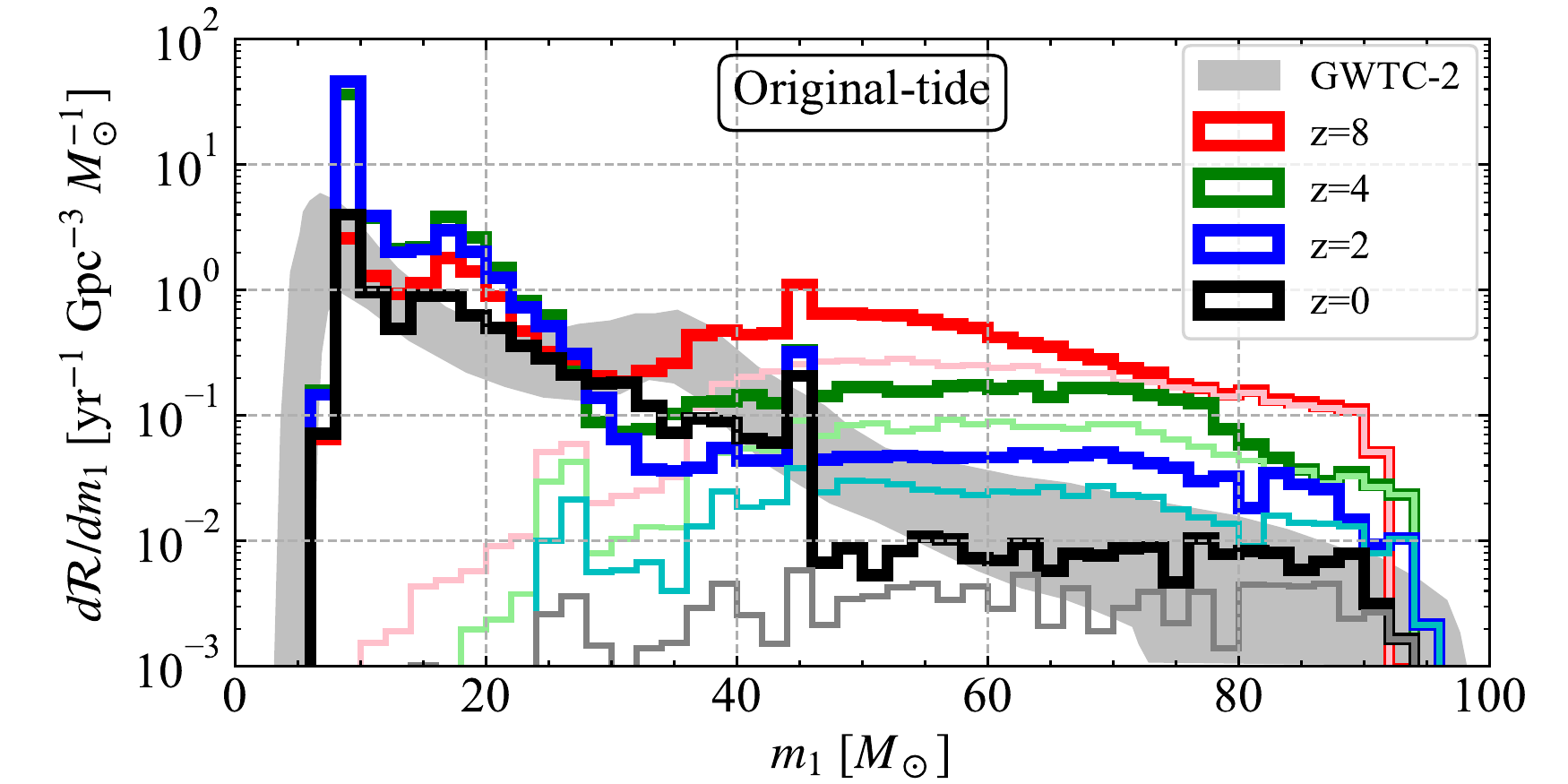}
  \plottwo{\fdir/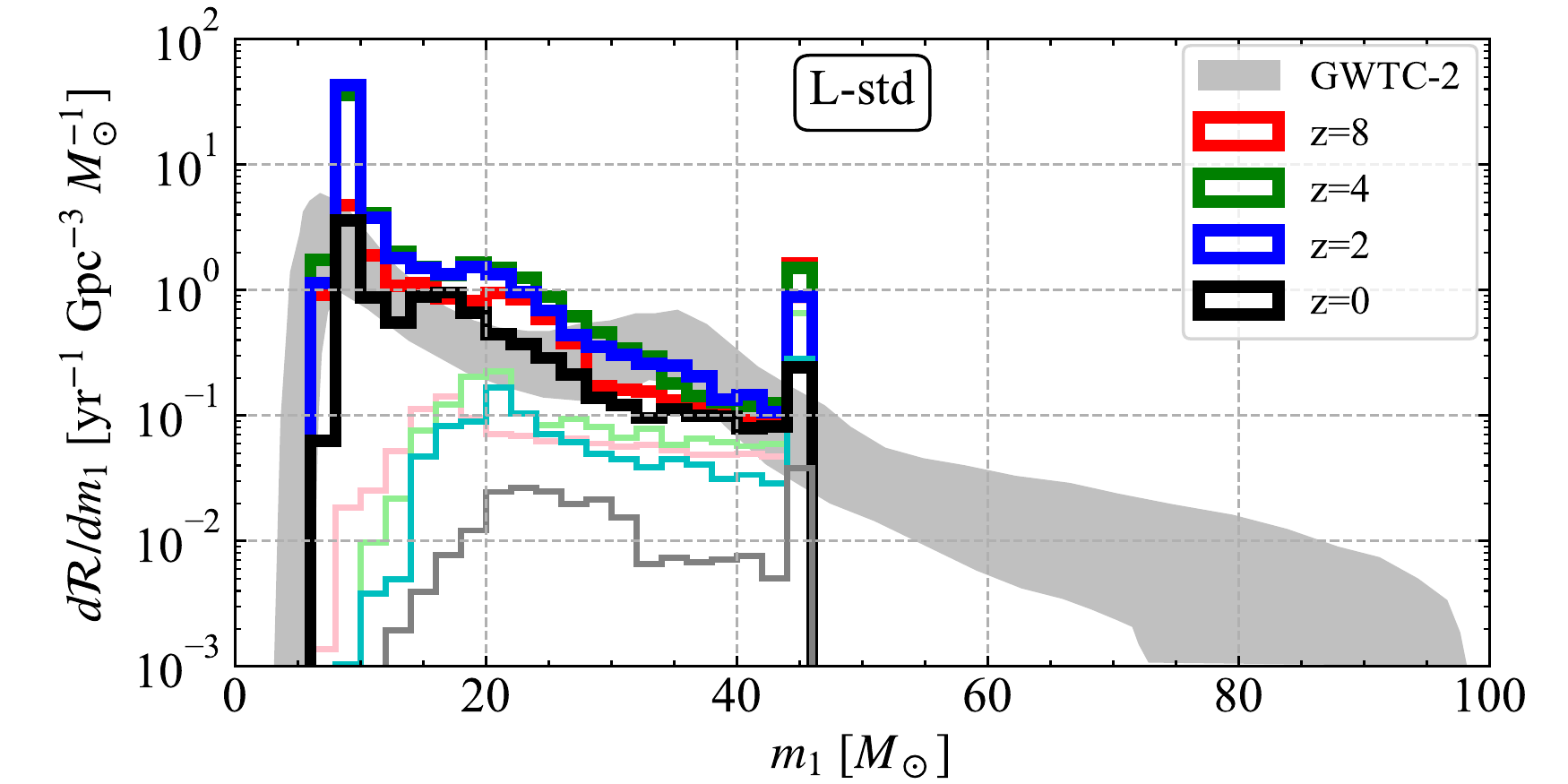}{\fdir/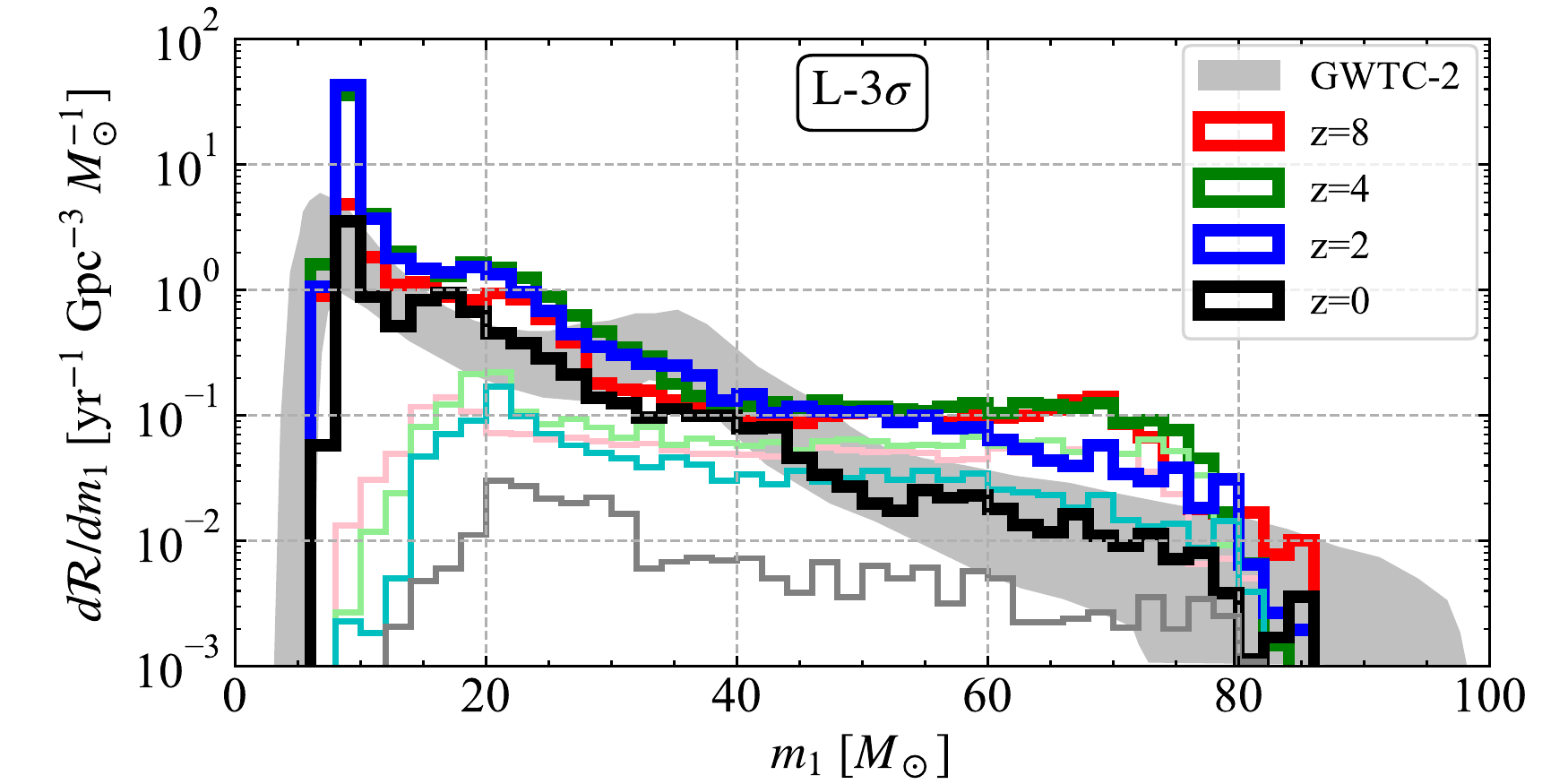}
  \caption{BH merger rate density differentiated by primary BH mass at
    the redshifts of $0$ (black), $2$ (blue), $4$ (green), and $8$
    (red) for parameter sets other than the fiducial one. The gray,
    cyan, light-green, and pink curves indicate Pop III contributions
    at the redshifts of $0$, $2$, $4$, and $8$, respectively. The
    shaded gray regions show the 90\% credible interval inferred by
    the LIGO-Virgo Gravitational-Wave Transient Catalog 2 (GWTC-2).}
  \label{fig:mergerRateMassRedshift}
\end{figure*}

Figure \ref{fig:mergerRateMassRedshift} shows the distributions of
primary BH masses at the redshifts of $z=0$, $2$, $4$, and $8$ for
parameter sets other than the fiducial one. We can see that in the
case of the top-light set (top-left) the BH merger rate for $m_1/\msun
\gtrsim 50$ at $z=0$ is definitely smaller than observed, although the
rate for $m_1/\msun \lesssim 50$ at $z=0$ is comparable to the
observation. For the stepwise set (top-right), the BH merger rate of
$m_1/\msun \gtrsim 50$ at $z=0$ is consistent with the observation,
although the PI mass gap rate is smaller than that in the fiducial
set. We also find that there are few binary BHs within the PI mass gap
in the case of the wide-binary set (middle-left). These results
demonstrate that the initial conditions of metal-poor binary stars
have great impacts on the formation of binary BHs within the PI mass
gap. Metal-poor binary stars should follow a top-heavy IMF and have
small initial pericenter distance in order to produce a sufficient
number of the PI mass gap events.

The stepwise set (top-right) shows an interesting result as for Pop
III binary stars. Pop III binary stars dominate the PI mass gap events
at any redshifts from $z=0$ to $8$. In other words, EMP binary stars
form little PI mass gap event. This is because all the
$Z/\zsun=10^{-4}$ stars have a top-light IMF, and the star formation
rate of $Z/\zsun=10^{-6}$ stars is small. If IMFs depend on
metallicity stepwisely like the stepwise set, the PI mass gap events
immediately indicate the presence of Pop III stars.

We next see the results of parameter sets with the L-model stars,
where the L-model stars experience efficient convective overshoot. The
result for the L-std set (bottom-left) shows that the primary BH mass
distribution for $m_1/\msun \lesssim 50$ is consistent with the
observed one, while that for $m_1/\msun \gtrsim 50$ is not. Even Pop
III and EMP binary stars fail to form binary BHs within the PI mass
gap. This result is consistent with \cite{2021ApJ...910...30T,
  2021MNRAS.505.2170T}. In the L model, Pop III stars with $\sim 80$
$\msun$ expand in radius up to a few $10^3$ $\rsun$. Their H envelope
is stripped through stable mass transfer or common envelope
evolution. Thus, they cannot form PI mass gap BHs through binary
evolution.

In contrast to the L-std set, the L-$\tsig$ set (bottom-right) can
produce binary BHs within the PI mass gap because of the different
criteria for PPI and PISN effects. The primary BH mass distribution at
$z=0$ is consistent with the observation, as is already pointed out by
\cite{2020ApJ...905L..15B}. The primary BH mass distribution in the
L-$\tsig$ set has no sudden drop at $m_1/\msun \sim 50$ unlike in the
fiducial set.  Recall that the sudden drop is caused by PPI effects
for the fiducial set. From this point of view, the fiducial set might
be slightly preferred over the L-$\tsig$ set.

\begin{figure}[ht!]
  \plotone{\fdir/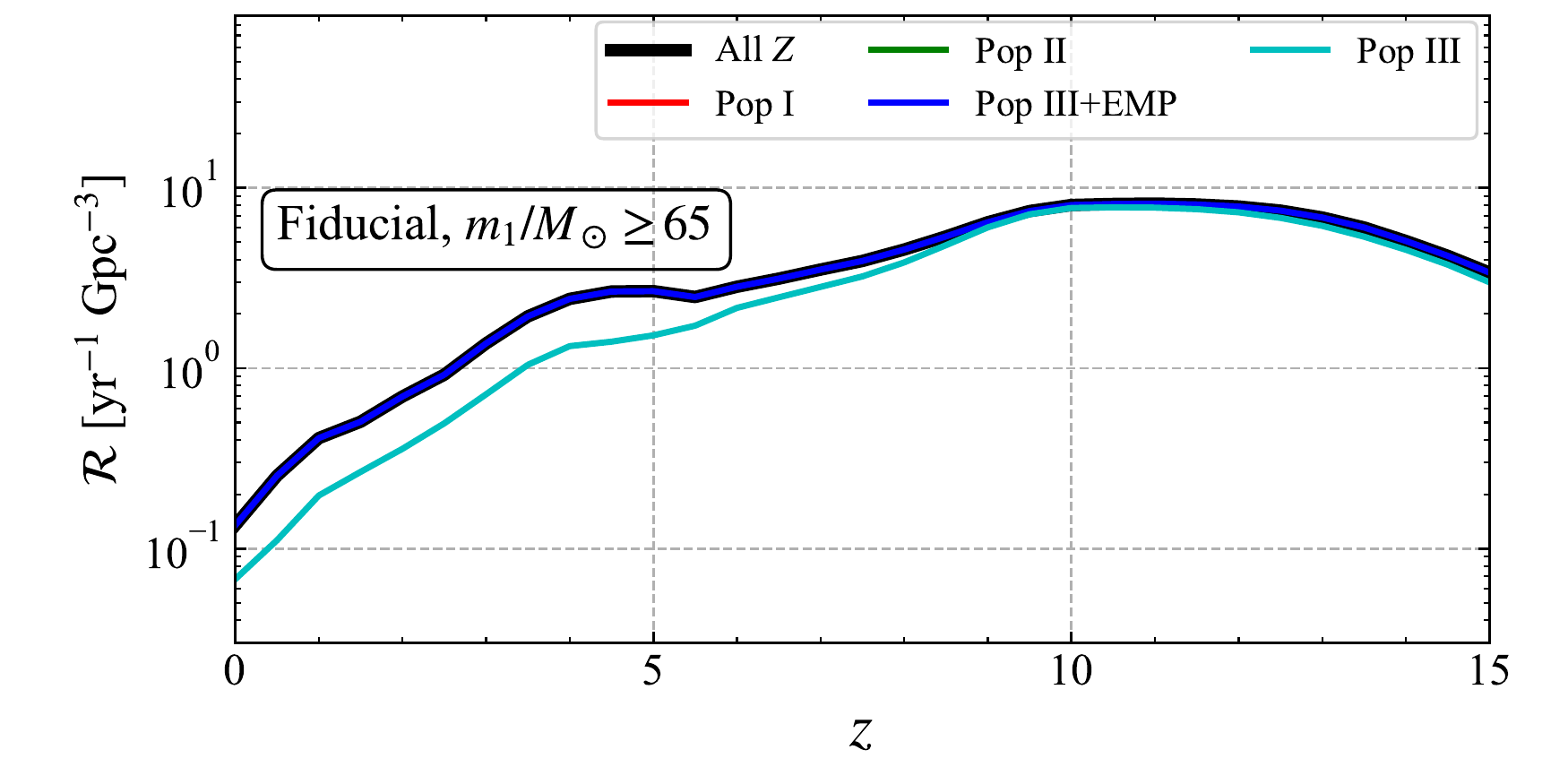}
  \plotone{\fdir/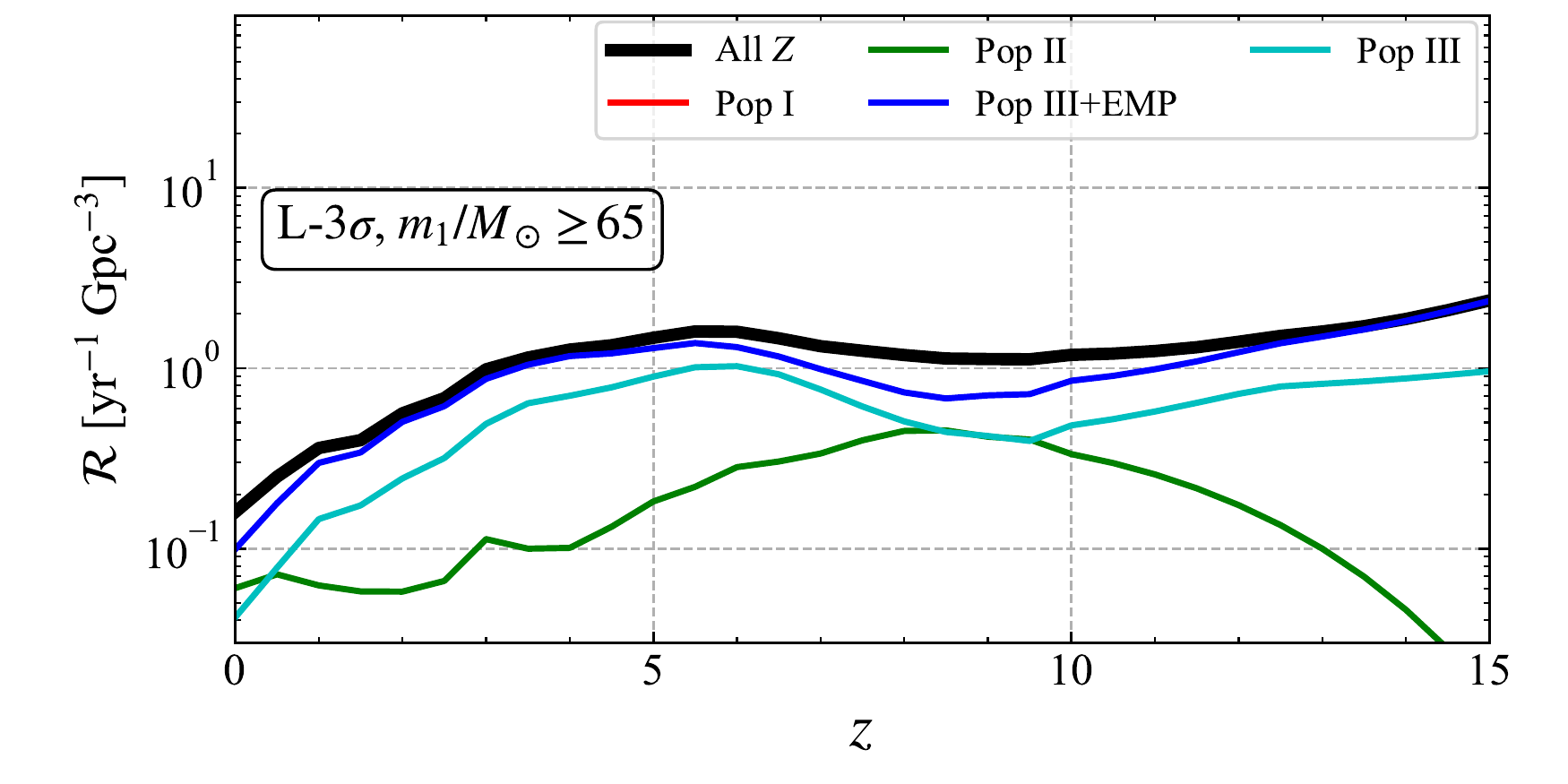}
  \caption{Redshift evolution of the BH merger rate density with
    $m_1/\msun \ge 65$ in the case with the fiducial parameter set
    (top) and the L-$\tsig$ set (bottom). The color codes are the same
    as in Figure \ref{fig:fdcl_mergerRateTotal}. Red curves in both
    sets, and green curve in the fiducial set can not be seen, since
    Pop I stars in both sets, and Pop II stars in the fiducial set do
    not form binary BHs with $m_1/\msun \ge 65$.}
  \label{fig:mergerRatePI}
\end{figure}

The striking difference between the fiducial and L-$\tsig$ sets is
seen in the redshift evolution of the BH merger rate density within
the PI mass gap (see Figure \ref{fig:fdcl_mergerRateMassRedshift} and
the bottom-left panel of Figure \ref{fig:mergerRateMassRedshift}). The
merger rate density increases monotonically from $z=0$ to $z=8$ for
the fiducial set, while it stops increasing at $z=4$ for the L-$\tsig$
set. This can be confirmed in Figure \ref{fig:mergerRatePI}. The
merger rate density has a peak at a redshift of $\sim 11$ in the
fiducial set, and at a redshift of $\sim 6$ in the L-$\tsig$ set. This
is because Pop III stars dominantly form the PI mass gap events in the
fiducial set, and do not in the L-$\tsig$ set. Note that Pop III SFR
has a peak at a redshift of $\sim 20$. There are two reasons why the
Pop III stars have smaller contribution to PI mass gap events in the
L-$\tsig$ set than in the fiducial set. First, Pop II stars can also
form PI mass gap events. Second, the number of Pop III stars forming
PI mass gap BHs becomes smaller in the L-$\tsig$ set than in the
fiducial set, since more massive Pop III stars form PI mass gap BHs in
the L-$\tsig$ set. In the fiducial set, Pop III stars with $65$--$90$
$\msun$ form the PI mass gap BHs, since they do not lose their H
envelopes through binary interactions, and collapse to BHs without
much mass loss. On the other hand, in the L-$\tsig$ set, Pop III stars
with $\gtrsim 130$ $\msun$ form the PI mass gap BHs. They expand to a
few $10^3$ $\rsun$, lose their H envelopes through binary
interactions, such as mass transfer or common envelope evolution, and
become totally naked He stars with $\gtrsim 65$ $\msun$. Finally, they
leave behind the PI mass gap BHs. Note that even stars with $65$--$90$
$\msun$ do not cause PPI nor PISN in the L-$\tsig$ set.
  
Surprisingly, we can find that Pop III binary stars have a significant
impact on the formation of the PI mass gap events at high redshift
even in the L-$\tsig$ set. As seen in the bottom panel of Figure
\ref{fig:mergerRateMassRedshift}, 1 of 2--3 PI mass gap events
originates from Pop III binary stars at $z \gtrsim 3$. Even if stars
experience efficient convective overshoot, Pop III binary stars can
produce sufficient the PI mass gap events, depending on the PI model.

The primary BH mass distribution in the original-tide set is similar
to that in the fiducial one. Pop III contributions become
  more important with increasing redshift, similarly to those in in
  the fiducial set. The prescription for tidal interaction has no
impact on the primary BH mass distribution.

\begin{figure*}[ht!]
  \plottwo{\fdir/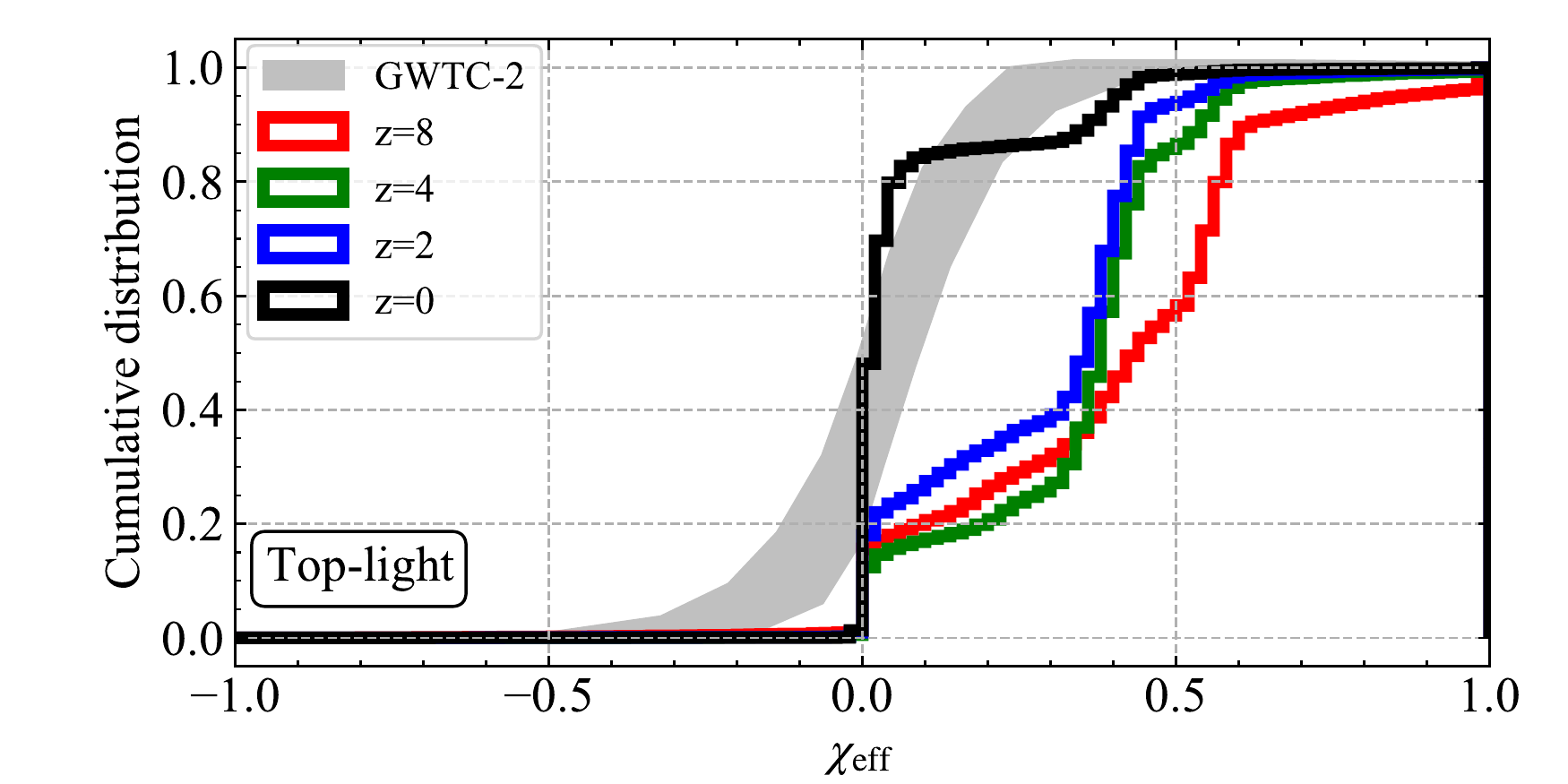}{\fdir/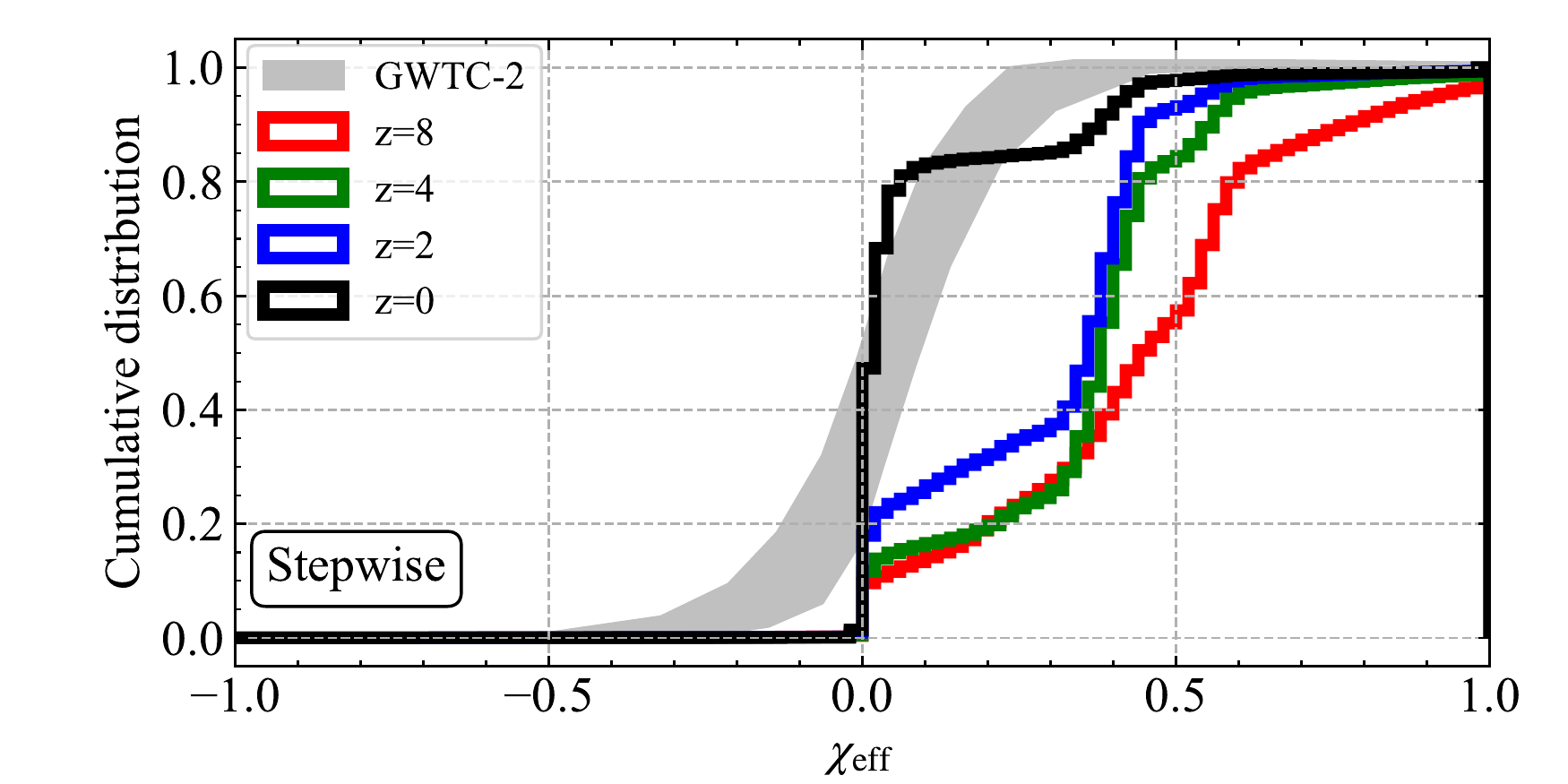}
  \plottwo{\fdir/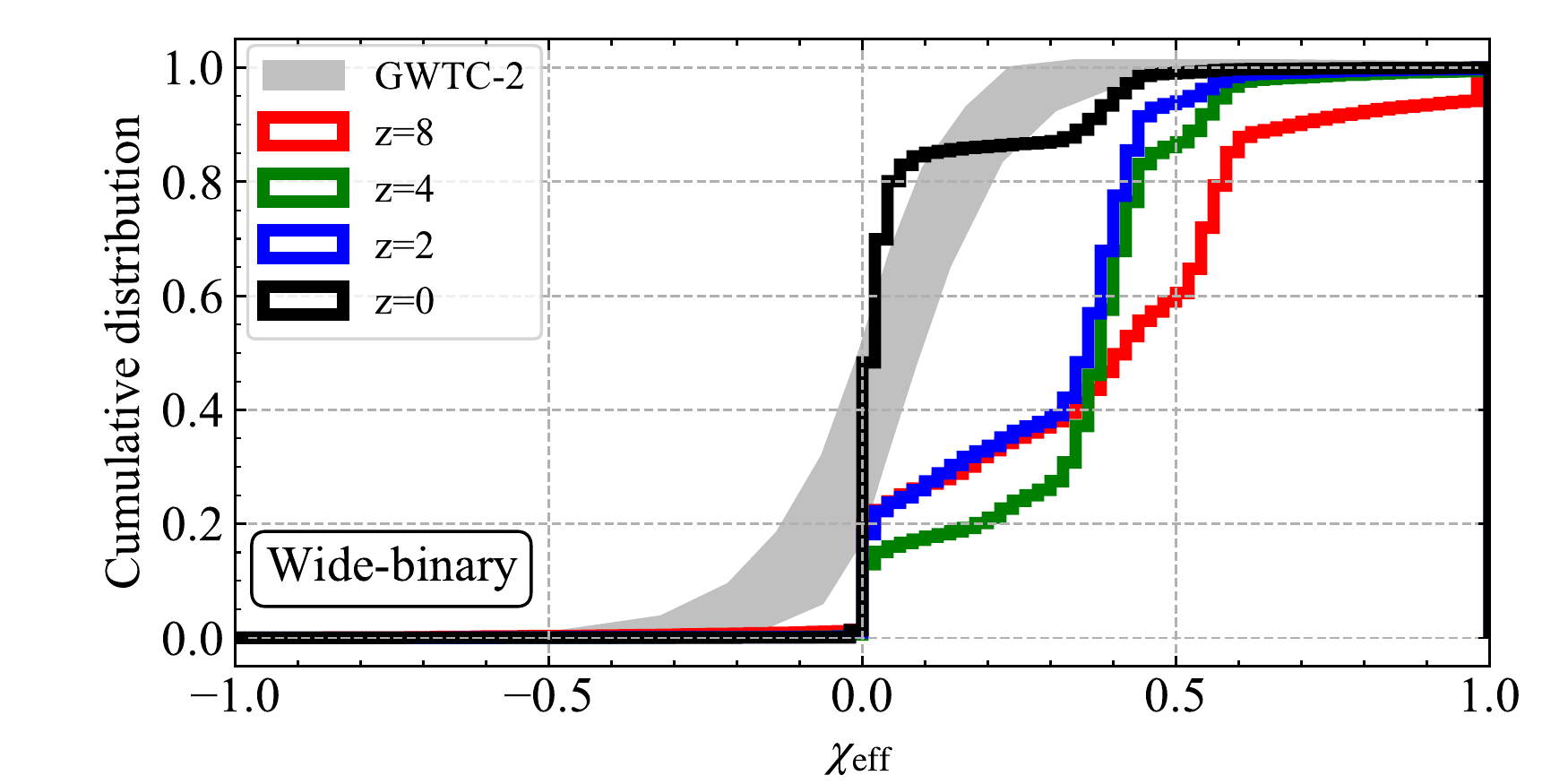}{\fdir/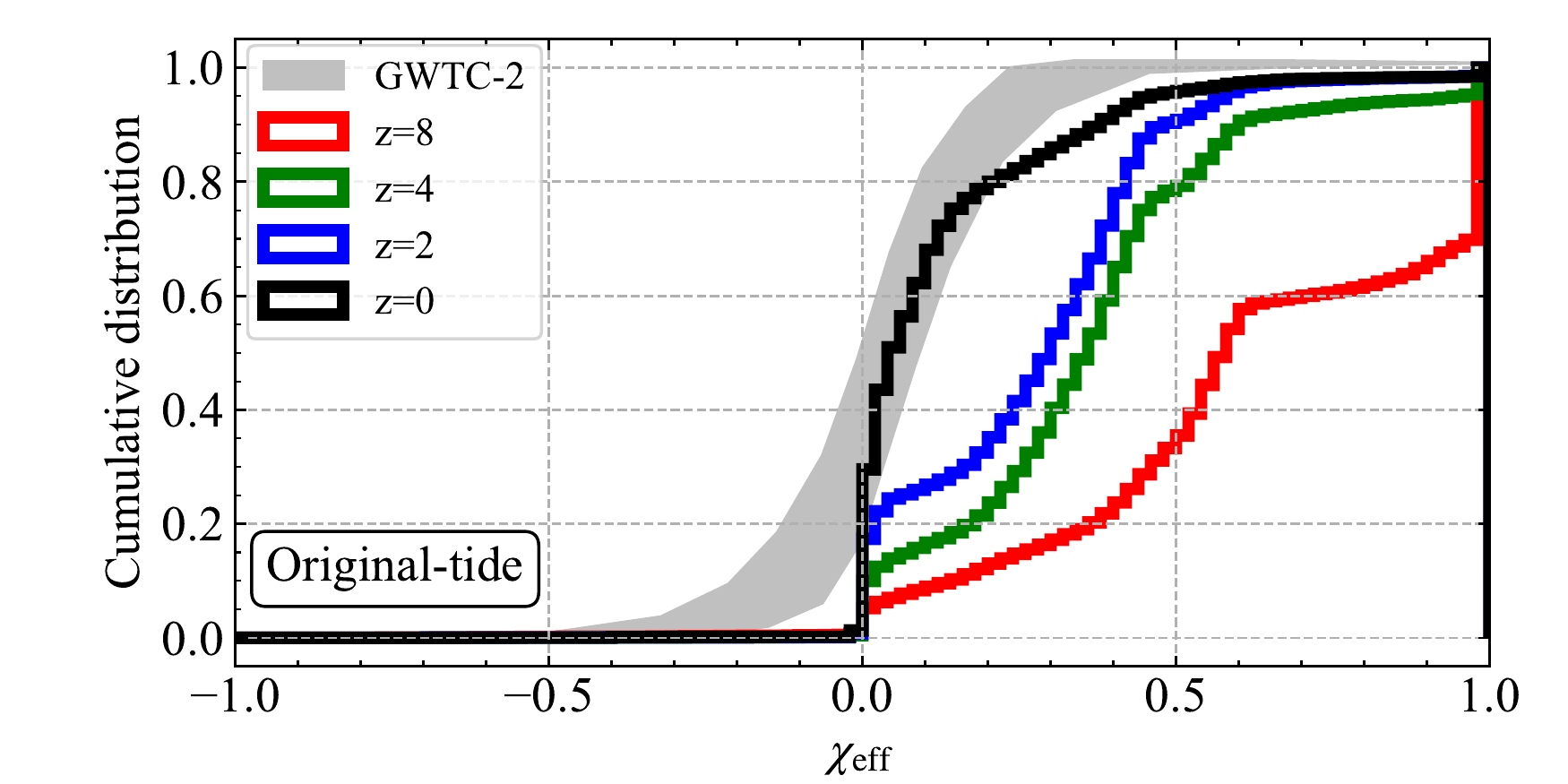}
  \plottwo{\fdir/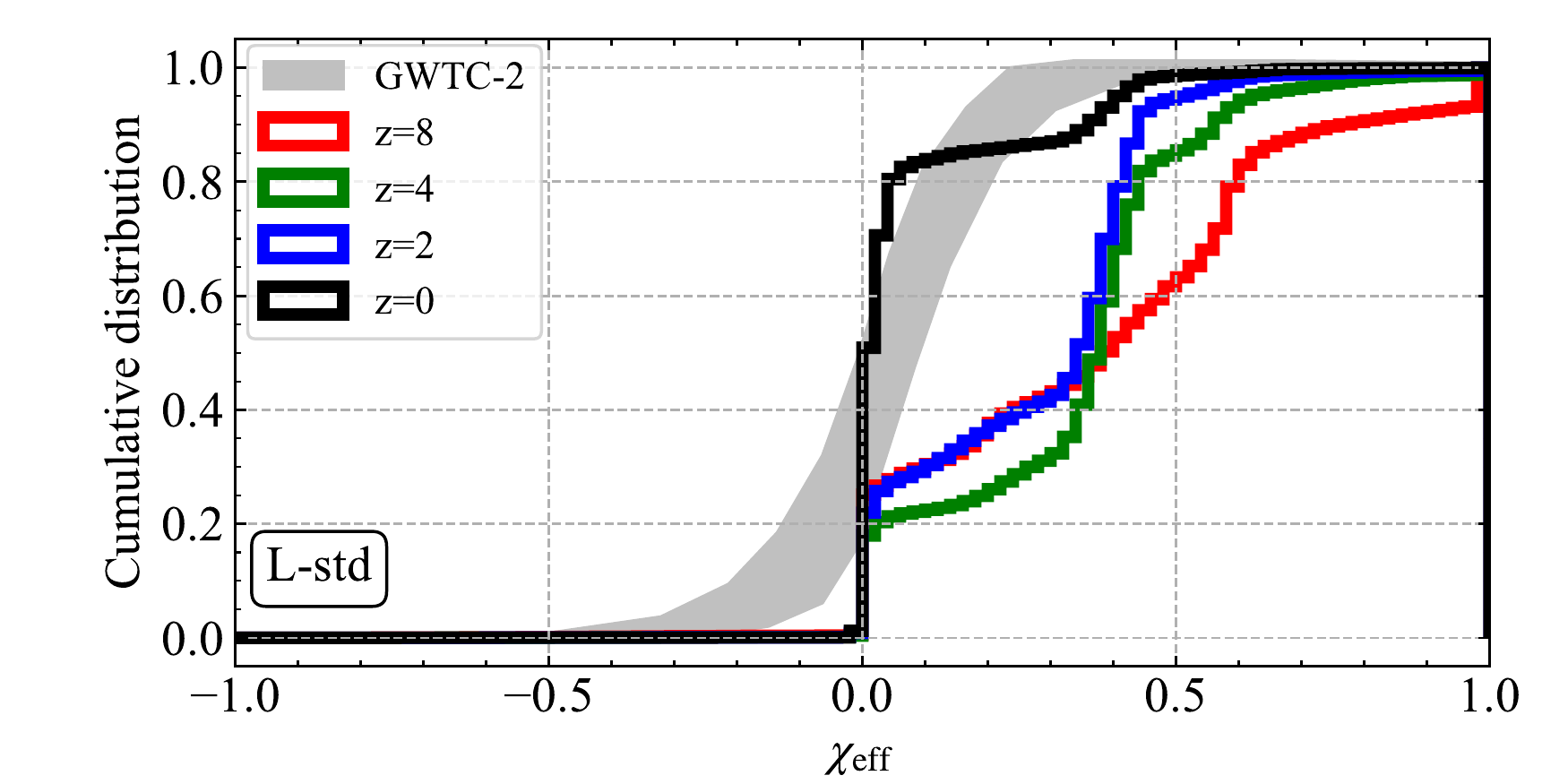}{\fdir/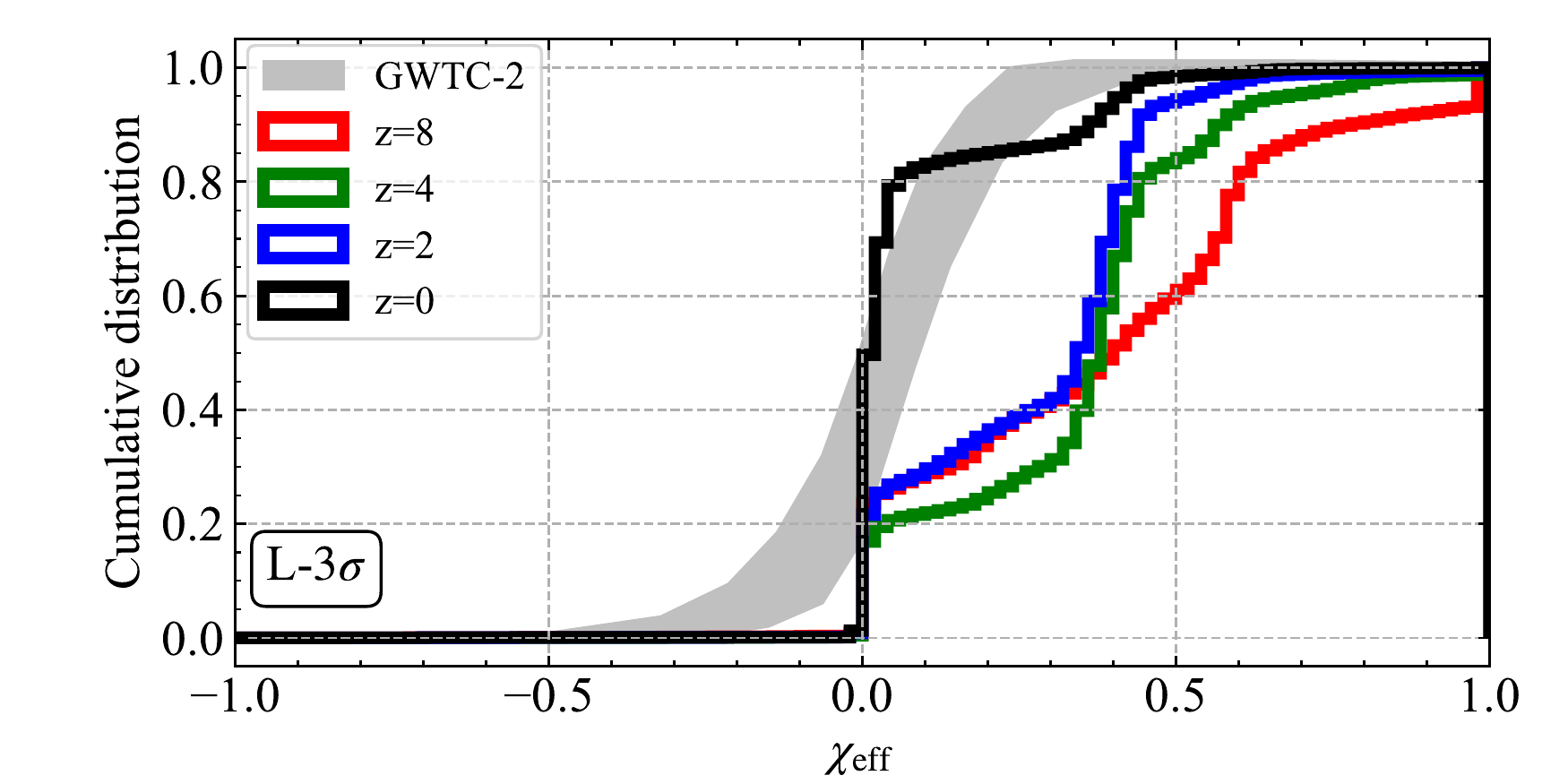}
  \caption{Cumulative distribution of effective spins for parameter
    sets other than the fiducial one.. The color codes are the same as
    Figure \ref{fig:mergerRateMassRedshift}, although Pop III
    contributions are not shown.}
  \label{fig:mergerRateXeffRedshift}
\end{figure*}

Figure \ref{fig:mergerRateXeffRedshift} shows the cumulative
distribution of effective spins for parameter sets other than the
fiducial one. Except for the original-tide set, the redshift evolution
of the cumulative $\xeff$ distributions is the same as for the
fiducial set. This is because merging binary BHs are dominated by
those with $m_1 \lesssim 50$ $\msun$ for all the sets. These binary BH
progenitors acquire their spin angular momenta in the same process for
all the sets: Pop I binary stars acquire spins at the phase of naked
He stars while all but Pop I binary stars are tidally spun up keeping
their H envelope.

The cumulative distribution in the original-tide set is different from
that in the fiducial set. In the original tidal prescription, naked He
stars are less spun up. Thus, binary BHs with $\xeff \sim 0.4$ in the
fiducial set are shifted to $\xeff \sim 0.2$--$0.4$ in the
original-tide set. In the original tidal prescription, blue supergiant
stars are more strongly spun up. Thus, binary BHs with $\xeff \gtrsim
0.5$ in the original-tide set contribute more than in the fiducial
one. About $50$ \% of binary BHs have $\xeff \sim 0$ in the fiducial
set, while few binary BHs have $\xeff \sim 0$ in the original-tide
set. This is because blue supergiant stars are spun up more as
described above.

In the original-tide set, the contribution of binary BHs with $\xeff
\gtrsim 0.5$ is too large to be reconciled with the observation. The
new tidal prescription seems to have advantages in reproducing the GW
observations, which show the dominance of binary BHs with small
$\xeff$ in GWTC-2 ($\sim 70$--$90$ \%) \citep{2021PhRvX..11b1053A,
  2021ApJ...913L...7A,2021ApJ...921L..15G,2021PhRvD.104h3010R}.

\section{Discussion}
\label{sec:Discussion}

As described in section \ref{sec:FiducialParameterSet}, the BH merger
rate density and its derivatives for the fiducial set are roughly
consistent with the results of GWTC-2 and GWTC-2.1. In this section,
we compare the results of the fiducial set with those of
previous studies.

First, we compare our results with two other isolated binary
scenarios, \cite{2020ApJ...905L..15B} and \cite{2021MNRAS.504L..28K},
who have claimed that isolated binary stars can reproduce the BH mass
distribution in the range of $5$--$100$ $\msun$ inferred by the GW
observations.  \cite{2020ApJ...905L..15B} have reported that Pop I/II
binary stars with metallicities $0.05$--$1.5$ $\zsun$ can make merging
binary BHs consistent with the GW observations with regard to the BH
mass and mass ratio distributions. Since we construct the $\tsig$ PI
model with reference to the model of \cite{2020ApJ...905L..15B}, the
results of the L-$\tsig$ set are naturally similar to those of
\cite{2020ApJ...905L..15B}. Here, we thus use the results of the
L-$\tsig$ set as a substitute for those of \cite{2020ApJ...905L..15B}
in discussing the differences between \cite{2020ApJ...905L..15B}'s and
our model predictions. As described in section
\ref{sec:OtherParameterSets}, the BH merger rate density with PI mass
gap BHs evolve differently with redshift between the models: it stops
increasing at $z \sim 6$ in the L-$\tsig$ set, while it is still
increasing up to $z \sim 11$ in the fiducial set. This is because Pop
III stars dominate the PI mass gap BHs in the fiducial set, and
partially form them in the L-$\tsig$ set. With future GW
observatories, such as Cosmic Explorer \citep{2019BAAS...51g..35R} and
Einstein Telescope \citep{2010CQGra..27s4002P}, which might be able to
detect merging binary BHs in this mass range up to $z \sim 10$, we can
pick out the right scenario.

\cite{2021MNRAS.504L..28K} have argued that merging binary BHs with
$m_1 \gtrsim 30\msun$ are dominated by Pop III binary stars. On the
other hand, not only Pop III but also Pop II and EMP binaries
dominantly yield merging binary BHs with $m_1 \gtrsim 50\msun$ in our
model. The future GW observatories can also distinguish these two
models because of their capability to detect binary BH mergers at high
redshifts. In the result of \cite{2021MNRAS.504L..28K}, Pop III binary
stars typically make binary BHs with $m_1 \sim 30\msun$. Thus, the BH
merger rate density with $m_1 \sim 30\msun$ should increase up to at
least $z \sim 8$. On the other hand, the BH merger rate density does
not change from $z=0$ to $8$ in our model (see Figure
\ref{fig:fdcl_mergerRateMassRedshift}).

In section \ref{sec:OtherParameterSets}, we show that Pop III and EMP
binary BHs strongly depend on their IMFs, period distributions, and
star evolution models. Here, we mention that Pop I/II binary BHs are
also sensitive to several uncertainties. For example, the following
two factors would make the local BH merger rate density much lower
than the observed one. First, the actual common envelope parameters
largely deviate from our choices, $\alpha=1$ and $\lambda \sim 0.1$
\citep{2012ApJ...759...52D}. Second, the metallicity distribution
($\sigmaz$) is much smaller than our choice, $\sigmaz = 0.35$
\citep{2021MNRAS.507.5224B}. In those cases, we need to consider
dynamical capture scenarios described below. In addition, the PI model
(see Eq. (\ref{eq:PImodel})) can affect the BH mass distribution. Even
if the $\cago$ reaction rate is fixed to the standard one, there are
several possibilities of the PPI effects
\citep{2019ApJ...882..121S}. The PPI effects can change the BH mass
distribution around $m_1 \sim 40$ $\msun$.

The observed primary BH mass distribution can also be explained by
dynamical capture scenarios in globular clusters
\citep[e.g.][]{2019PhRvD.100d3027R}, open clusters
\citep[e.g.][]{2020MNRAS.495.4268K, 2020MNRAS.497.1043D,
  2020ApJ...898..152S}, or active galactic nuclei (AGN) disks
\citep[e.g][]{2020ApJ...898...25T}. The difference between those
models and ours is as follows.  While binary BHs with $m_1 \sim
100-130\msun$ are in principle allowed to form in the dynamical
capture scenarios, they are not in our scenario. Note that binary BHs
with $m_1 \gtrsim 130\msun$ can form if $\gtrsim 300$ $\msun$ stars
are present \citep{2019ApJ...883L..27M, 2021ApJ...910...30T,
  2021MNRAS.505L..69H}, since they overcome PPI and PISNe. Thus, the
primary BH mass distribution should smoothly extend beyond $m_1 \sim
100\msun$ if the dynamical capture is the dominant formation process
of merging binary BHs.  If our scenario is correct, the BH merger rate
density should suddenly decrease at $m_1 \sim 100-130\msun$. The
fourth and fifth observing runs of the current GW observatories
possibly assess the validity of these two scenarios if these observing
runs detect binary BHs with $m_1 \sim 100-130\msun$, or put some
constraints on the upper limit of a BH merger rate in this $m_1$
range.

\section{Summary}
\label{sec:Summary}

We have derived the BH merger rate density and the distributions of
merger properties by means of BPS calculations for $Z/\zsun = 0$--$1$
binary stars equivalent to Pop I/II/III and EMP binary stars, which
span an unprecedentedly wide metallicity range. We adopted various
initial conditions of binary stars and single and binary star
models. With the fiducial set of the parameters, we successfully
obtained the BH merger rate density, the primary BH mass in the range
$5$--$100$ $\msun$, and effective spin distributions consistent with
the current GW observations. In our results, Pop III and EMP binary
stars are responsible for the PI mass gap events like
GW190521. Moreover, Pop III binary stars become more responsible for
the PI mass gap events with increasing redshift, and contribute most
of them at $z \gtrsim 8$.  We have found that Pop III and EMP binary
stars should have a top-heavy IMF, small pericenter distance at the
initial time, and inefficient convective overshoot in order to produce
a sufficient number of the PI mass gap events to explain the PI mass
gap event rate inferred by GW observations.

We examine the dependence of binary BH properties on Pop III (and EMP)
IMFs, period distributions, and star evolution models with different
convective overshoot. We note that Pop III stars also contain other
uncertain parameters, such as Pop III SFR and binary fraction, since
Pop III stars have not yet been discovered. These uncertainties would
directly affect our predicted BH merger rate density, especially the
PI mass gap event rate mainly yielded by Pop III (and EMP) binary
stars in our model. Apart from Pop III and EMP stars, Pop I/II binary
BHs can also depend on initial conditions (SFR and metallicity
distribution) and binary evolution parameters (common envelope
parameters and mass transfer parameters). If they largely deviate from
our choice, the local BH merger rate density can be much smaller (or
larger) than the observed one. However, examining the dependence is
beyond of this paper.

Finally, we emphasize the Pop III contributions to the formation of
the PI mass gap events. In the parameter sets which reproduce the
observed primary BH mass distribution (the fiducial, stepwise,
original-tide, and L-$\tsig$ sets), Pop III binary stars have
important roles in forming the PI mass gap events. In the fiducial,
stepwise, and original-tide sets, Pop III binary stars dominate the PI
mass gap at $z \gtrsim 8$. Even in the L-$\tsig$ set, 1 of 2--3 PI
mass gap events originates from Pop III binary stars at $z \gtrsim
3$. This means that future GW observations of the PI mass gap events
will be very useful for Pop III studies, if one of the above isolated
binary scenarios \citep[][and this
  paper]{2020ApJ...905L..15B,2021MNRAS.504L..28K} is correct.

\begin{acknowledgments}
  The authors appreciate the anonymous referee for the thorough
  reading and many fruitful suggestions. The authors wish to express
  their cordial gratitude to Prof. Takahiro Tanaka, general PI of
  Innovative Area Grants-in-Aid for Scientific Research "Gravitational
  wave physics and astronomy: Genesis" for his continuous interest and
  encouragement. This research could not been accomplished without the
  support by Grants-in-Aid for Scientific Research (17H01101,
  17H01102, 17H01130, 17H02869, 17H06360, 17K05380, 19K03907,
  19H01934, 20H00158, 20H05249, 21K13914) from the Japan Society for
  the Promotion of Science. This research has made use of data,
  software and/or web tools obtained from the Gravitational Wave Open
  Science Center (\url{https://www.gw-openscience.org/}), a service of
  LIGO Laboratory, the LIGO Scientific Collaboration and the Virgo
  Collaboration. LIGO Laboratory and Advanced LIGO are funded by the
  United States National Science Foundation (NSF) as well as the
  Science and Technology Facilities Council (STFC) of the United
  Kingdom, the Max-Planck-Society (MPS), and the State of
  Niedersachsen/Germany for support of the construction of Advanced
  LIGO and construction and operation of the GEO600
  detector. Additional support for Advanced LIGO was provided by the
  Australian Research Council. Virgo is funded, through the European
  Gravitational Observatory (EGO), by the French Centre National de
  Recherche Scientifique (CNRS), the Italian Istituto Nazionale di
  Fisica Nucleare (INFN) and the Dutch Nikhef, with contributions by
  institutions from Belgium, Germany, Greece, Hungary, Ireland, Japan,
  Monaco, Poland, Portugal, Spain.
\end{acknowledgments}

\software{Open data from the first and second observing runs of
  Advanced LIGO and Advanced Virgo \citep{2021SoftX..1300658A}; {\tt
    HOSHI} \citep{2016MNRAS.456.1320T, 2018ApJ...857..111T,
    Takahashi19, Yoshida19}; {\tt BSE} \citep{2000MNRAS.315..543H,
    2002MNRAS.329..897H, 2020MNRAS.495.4170T, 2021MNRAS.505.2170T};
  {\tt Matplotlib} \citep{2007CSE.....9...90H}; {\tt NumPy}
  \citep{2011CSE....13b..22V}.}

\appendix

\section{Single star models}
\label{sec:SingleStarModels}

In this section, we show the evolutions of the M- and L-model stars
with various metallicities, and compare them with those of the
SSE-model stars as a reference. We remark again that the M- and
L-model stars have inefficient and efficient convective overshoot,
respectively. Before showing their evolutions, we overview how to
construct evolution tracks of the M- and L-model stars, which is
described in detail in \cite{2020MNRAS.495.4170T}. We perform 1D
hydrodynamics simulations of stars with various metallicities and
masses. Based on the simulation results, we make fitting formulae of
stellar evolution tracks, and incorporate the fitting formulae into
the BSE code. The main components of the 1D hydrodynamics simulations
are as follows. We use a 1D stellar evolution code, \texttt{HOSHI}
code \citep{2016MNRAS.456.1320T, 2018ApJ...857..111T, Takahashi19,
  Yoshida19}. We take into account convection, semiconvection, and
convective overshoot for chemical mixing, and do not account for
rotation and rotational mixing. We input stellar physics with nuclear
reaction network with 49 species of nuclei
\citep{2018ApJ...857..111T}, stellar equation of state
\citep{Blinnikov96, Vardya60, Iben63}, OPAL, molecular, and conductive
opacities \citep[][respectively]{Iglesias96, Ferguson05, Cassisi07},
and neutrino energy loss \citep{Itoh96}.

The M- and L-model stars are modeled as stars with the less and more
efficient convective overshoot, respectively. In the \texttt{HOSHI}
code, we treat convective overshoot as a diffusive process above
convective regions. The diffusion coefficient of the convective
overshoot exponentially decreases with the distance from the
convective boundary as
\begin{align}
  D_{\rm cv}^{\rm ov} = D_{\rm cv,0} \exp \left( -2 \frac{\Delta
    r}{f_{\rm ov} H_{\rm P0}} \right),
\end{align}
where $D_{\rm cv,0}$ and $H_{\rm P0}$ are the diffusion coefficient
and the pressure scale height at the convective boundary,
respectively, and $\Delta r$ is the distance from the convective
boundary. The overshoot parameter $f_{\rm ov}$ is set to be $0.01$ for
the M-model stars, and $0.03$ for the L-model stars, which are the
same values of Sets M$_{\rm A}$ and L$_{\rm A}$ in \cite{Yoshida19},
respectively. The overshoot parameter of the M-model stars is adjust
to the MS width of AB stars observed in open clusters in the Milky Way
Galaxy \citep{Ekstroem12}.  On the other hand, the overshoot parameter
of the L-model stars are determined based on the calculation to
early-B type stars in the Large Magellanic Cloud similarly to Stern
model \citep{2011A&A...530A.115B}.

\begin{figure*}[ht!]
  \plotone{\fdir/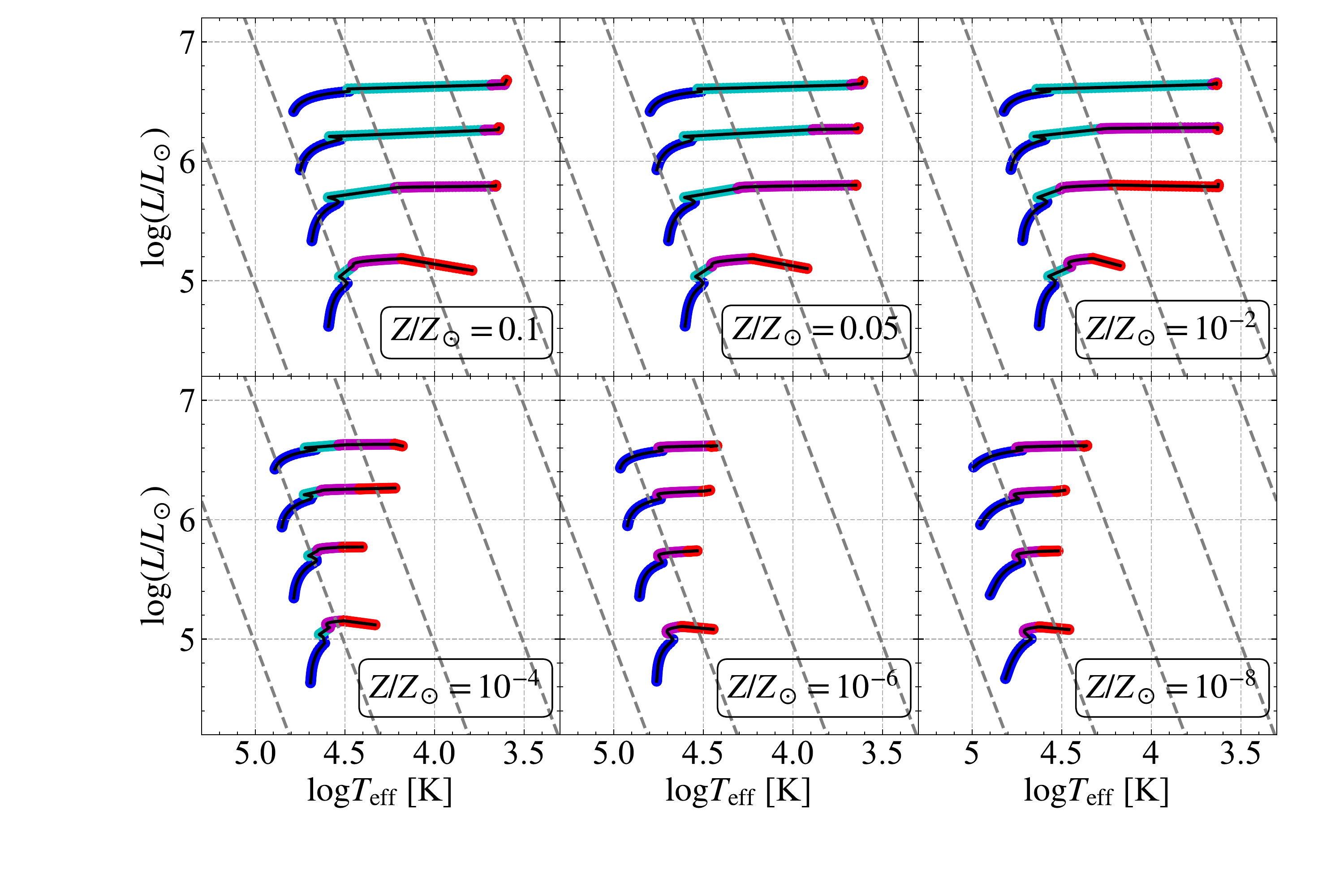}
  \caption{Hertzsprung-Russell diagram of the M model (our inefficient
    convective overshoot model). Each metallicity is indicated in each
    panel. The stellar masses are $20$, $40$, $80$, and $160$ $\msun$
    from bottom to top in each panel. Colors show stellar phases: main
    sequence (blue), Hertzsprung gap (cyan), core helium burning
    (magenta), and shell helium burning (red) phases. The gray dashed
    lines show stellar radii of $1$, $10$, $10^2$, $10^3$, and $10^4$
    $\rsun$ from left to right. Stellar wind mass loss is not
    considered for their evolution.}
  \label{fig:hrdSingleMmodel}
\end{figure*}

\begin{figure*}[ht!]
  \plotone{\fdir/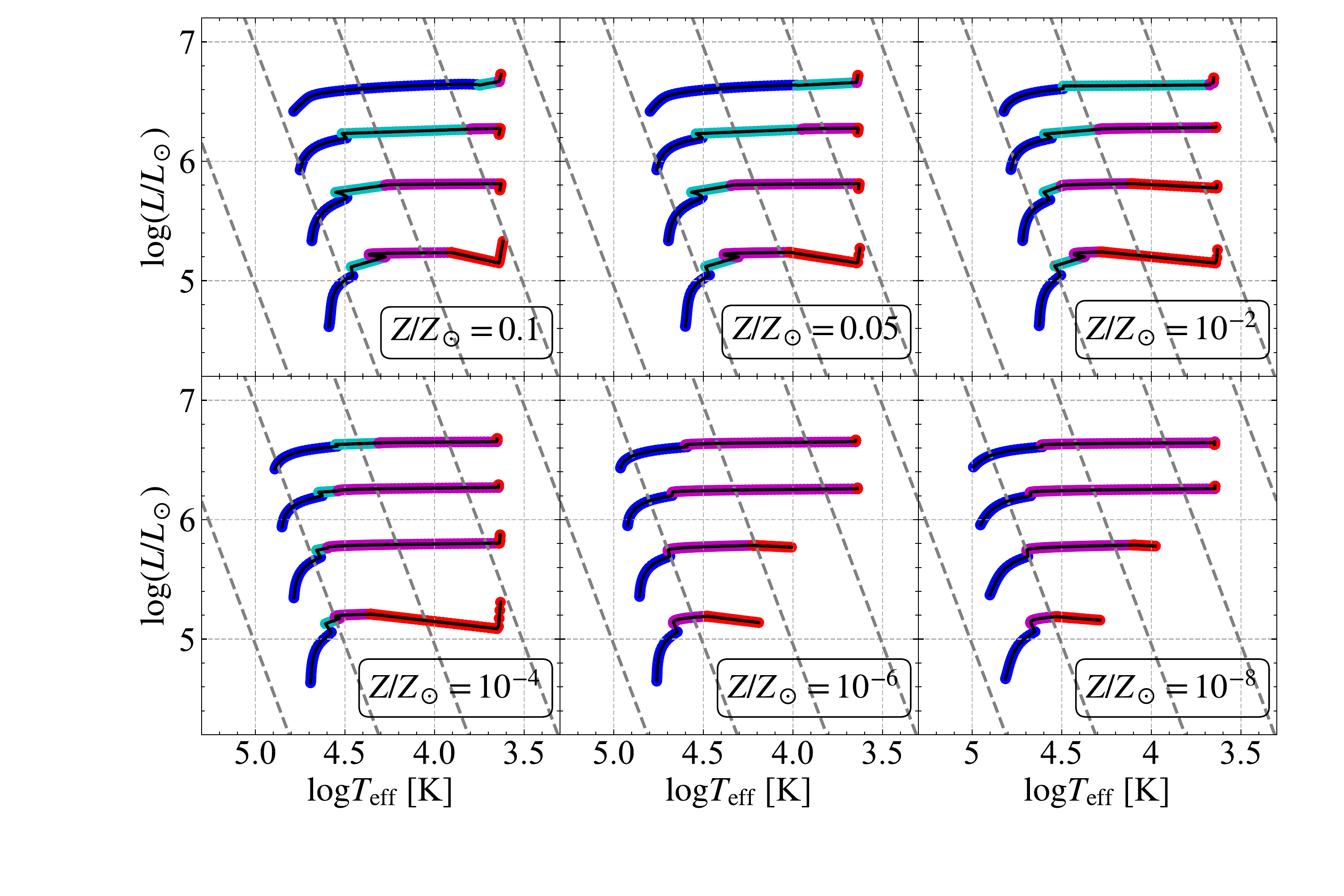}
  \caption{The same as Figure \ref{fig:hrdSingleMmodel}, except for
    the L model (our efficient convective overshoot model).}
  \label{fig:hrdSingleLmodel}
\end{figure*}

Figures \ref{fig:hrdSingleMmodel} and \ref{fig:hrdSingleLmodel} and
show the Hertzsprung-Russell (HR) diagrams of the M-model and L-model
stars, respectively. We do not account for stellar wind mass loss when
making these stellar evolution tracks.

The most striking difference between the M- and L-model stars is the
radius evolution of Pop III and EMP stars with $m/\msun =
80$--$160$. The L-model stars expand to more than $10^3$ $\rsun$ and
evolve to red supergiant stars, while the M-model stars expand up to
$\sim 10^2$ $\rsun$ and keep blue supergiant stars. Since the L-model
stars involve more efficient convective overshoot, they transfer large
amounts of H in their radiative envelopes into their convective cores,
and form large He cores at the end of their MS phases. For example,
the L-model stars with $m/\msun = 80$ and $Z/\zsun = 10^{-8}$ (or Pop
III) leave a $42$ $\msun$ He core. Such large He cores create large
luminosities in their post-MS phases, and expand their H envelopes. On
the other hand, the M-model stars form small He cores in their MS
phases. As a reference, the M-model stars with $m/\msun = 80$ and
$Z/\zsun = 10^{-8}$ (or Pop III) leave a $36$ $\msun$ He core. Since
such small He cores emit small luminosities, and their H envelopes
have small opacities because of their low metallicities, they do not
expand much. \cite{2021MNRAS.504..146V} have also found similar
results.

Both of the M- and L-model stars evolve to red supergiant stars if
they are Pop I/II stars with $m/\msun = 40$--$160$. This is because
their He cores are large and the opacities of their H envelopes are
large due to their high metallicities. Both of the M- and L-model
stars tend to keep blue supergiant stars if they are Pop III and EMP
stars with $m/\msun = 20$--$40$. Their luminosities are not large
enough to expand their radii to red supergiant stars.

\begin{figure*}[ht!]
  \plotone{\fdir/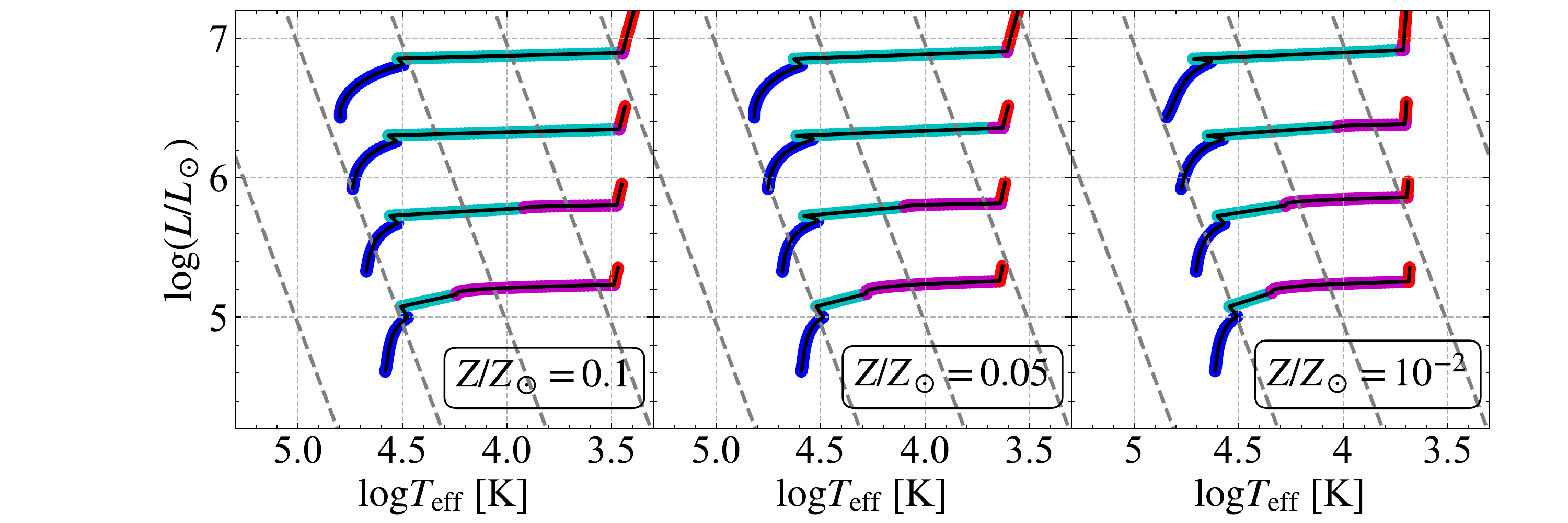}
  \caption{The same as Figure \ref{fig:hrdSingleMmodel}, except for
    the SSE model. Note that the SSE model does not support stars with
    $Z/\zsun = 10^{-4}, 10^{-6}$, and $10^{-8}$. }
  \label{fig:hrdSingleSsemodel}
\end{figure*}

Figure \ref{fig:hrdSingleSsemodel} shows the HR diagrams of the
SSE-model stars as a reference. Note that the SSE model does not
prepare stars with $Z/\zsun < 5 \times 10^{-3}$. They expand to more
than $10^3$ $\rsun$ for all masses and $Z$. These features are similar
to the L-model stars.

\section{Comparison between the SSE- and M-model stars}
\label{sec:ComparisonWithSse}

In this section, we compare between the BH merger rate densities in
the fiducial and SSE-std sets as a reference. Additionally, we also
make comparison between the L-std and SSE-std sets. The difference
between these sets is single star models with Pop II
metallicities. Thus, we focus only on their results.

\begin{figure}[ht!]
  \plottwo{\fdir/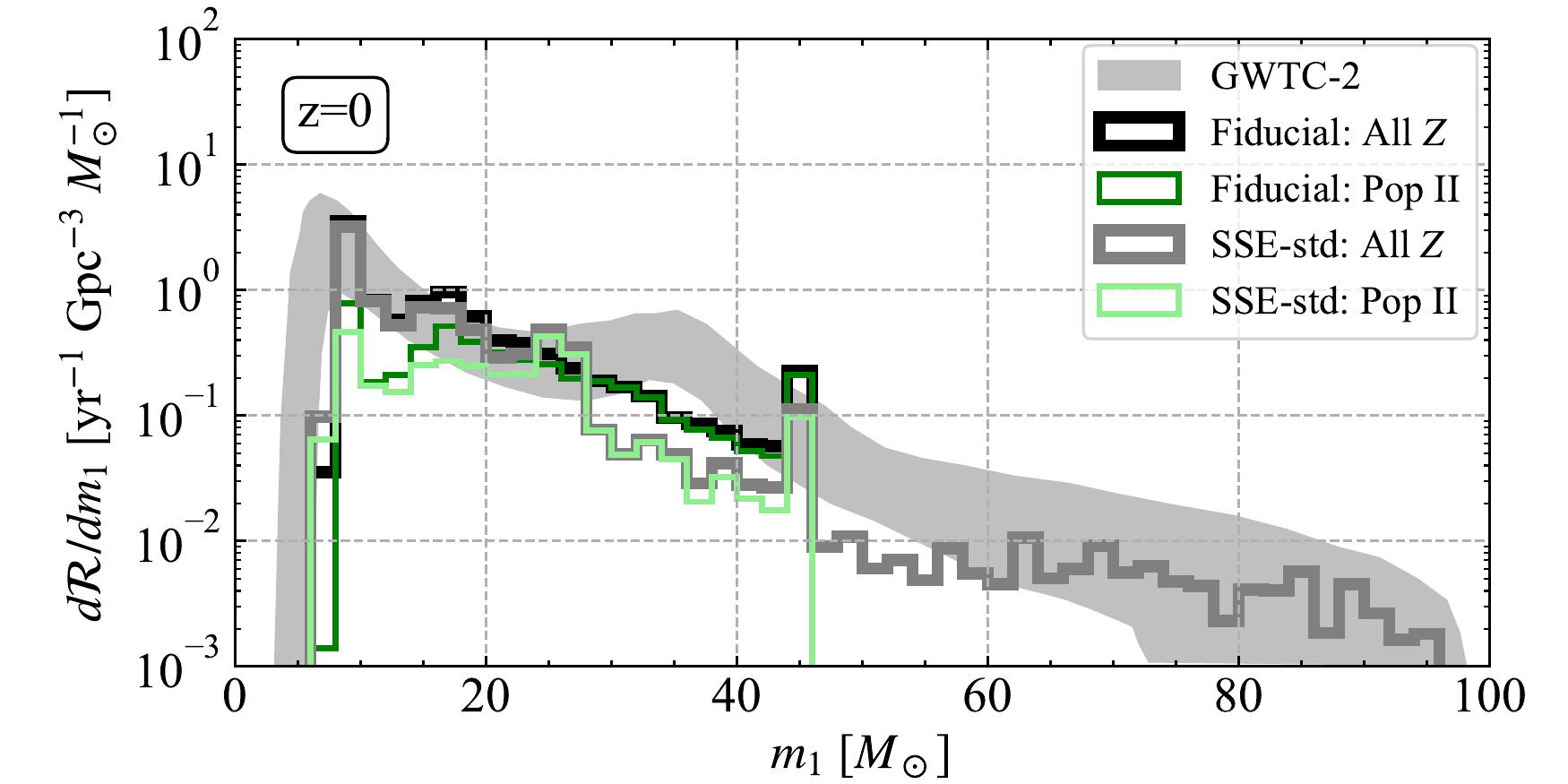}{\fdir/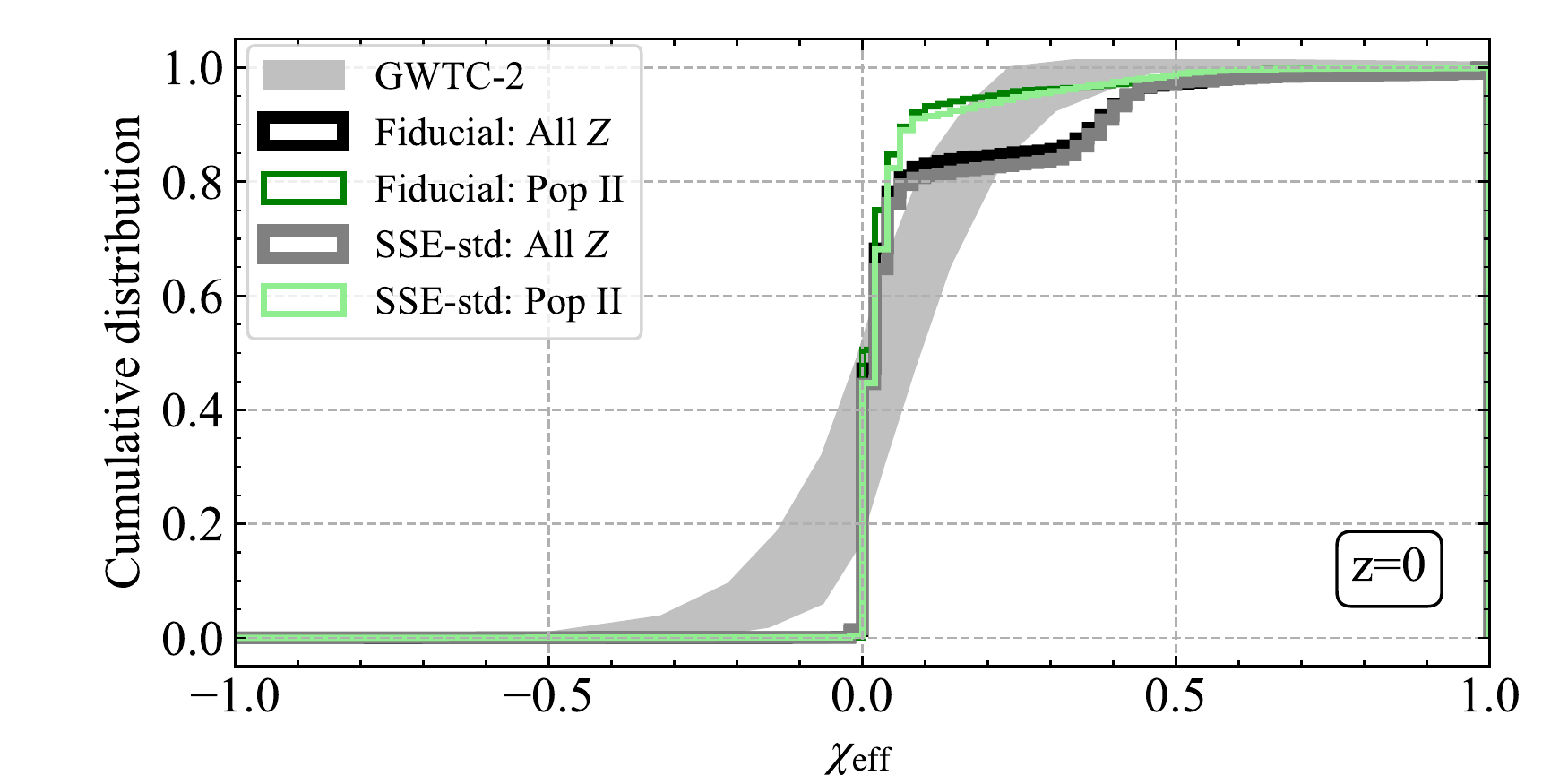}
  \plottwo{\fdir/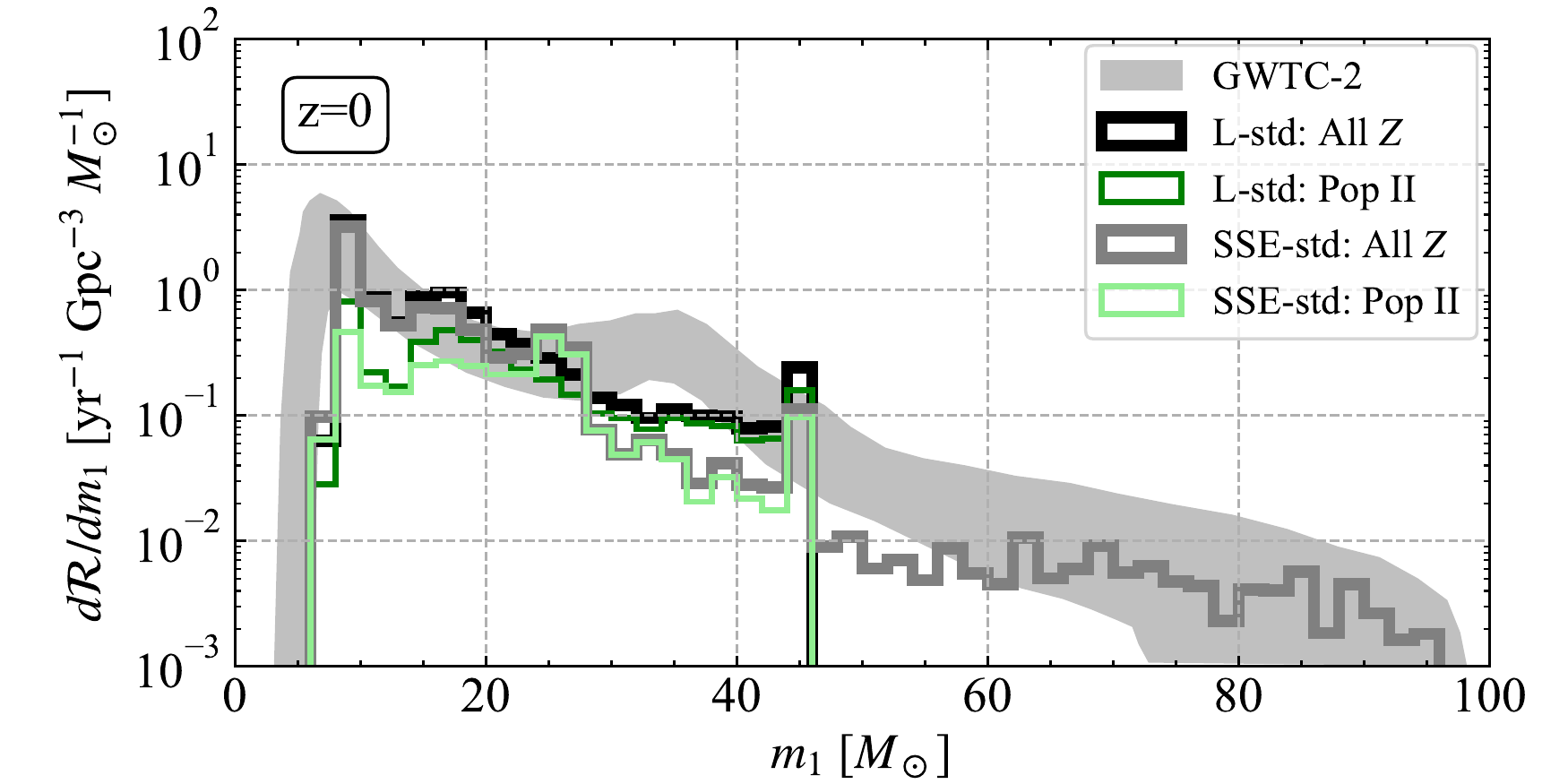}{\fdir/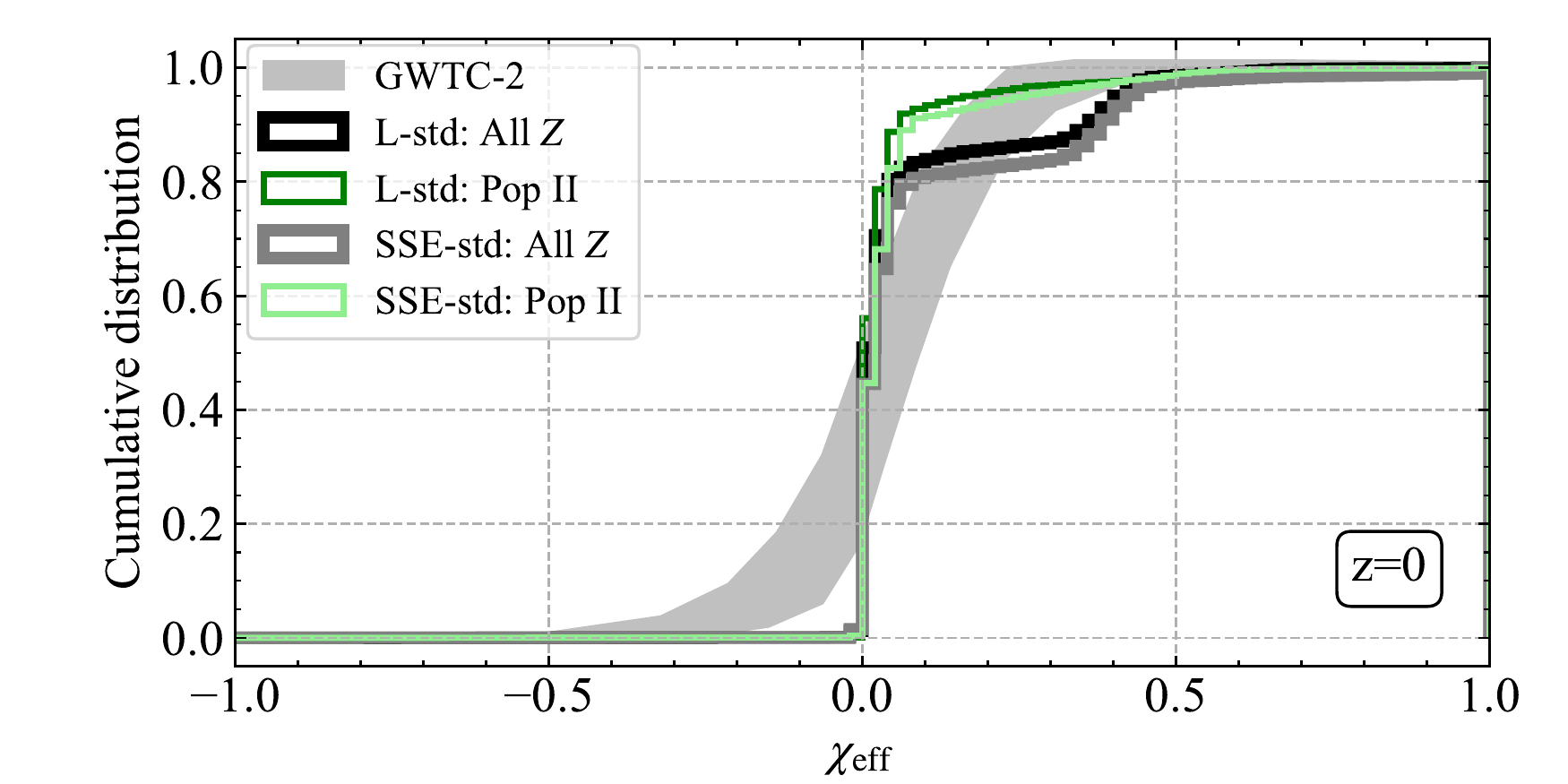}
  \caption{BH merger rate density differentiated by primary BH mass
    and cumulative distribution of effective spins at the redshift of
    $z=0$ for the fiducial and SSE-std sets (the top panels) and for
    the L-std and SSE-std sets (the bottom panels). The color codes of
    the fiducial and L-std sets are the same as Figures
    \ref{fig:fdcl_mergerRateMassMetal} and
    \ref{fig:fdcl_mergerRateXeffMetal}. The color codes of the SSE-std
    set are lighter than the counterparts of the fiducial set.}
  \label{fig:comparisonBse}
\end{figure}

The top panels of Figure \ref{fig:comparisonBse} show the mass and
spin distributions of merging binary BHs at the redshift of $z=0$ in
the fiducial and SSE-std sets. Roughly speaking, the mass and spin
distributions in the SSE-std set are in good agreement with those in
the fiducial set.

Especially, their spin distributions are quite similar. So, we focus
only on their mass distributions. The primary BH masses in the SSE-std
set are cut off at $m_1/\msun = 45$ because of PPI and PISN effects,
similarly to those in the fiducial set. The SSE-model stars expand to
more than $10^3$ $\rsun$ and evolve to red supergiant stars. They lose
their H envelopes through stable mass transfer or common envelope
evolution. Thus, they cannot form PI mass gap BHs.

The BH merger rate density in the SSE-std set is a few times smaller
than in the fiducial set in the range of $m_1/\msun \sim
30$--$40$. The reason is as follows. The SSE-model stars have larger
radii in their MS phases than the M-model stars. Thus, the SSE-model
stars lose more masses in their MS phases through stable mass
transfer. Then, they leave smaller He cores. In their post-MS phases,
they further lose their masses through stable mass transfer, and
finally leave naked He stars. Note that they have companion stars
close enough to merge within the Hubble time. Thus, the resulting BHs
in the SSE-std set have smaller mass than in the fiducial set, and the
BH merger rate density with $m_1/\msun \sim 30$--$40$ is smaller.

In summary, the M-model and SSE-std stars generate consistent binary
BHs, although there are several small differences. Moreover, when it
comes to Pop II results, the M/L-model and SSE-std stars yield similar
binary BHs.

\section{Effects of the metallicity choice in SSE-model stars}
\label{sec:EffectsOfBinning}

For the SSE-std set, we choose SSE-model stars with metallicities
$0.25, 0.5$, and $1\zsun$ for Pop I stars, SSE-model stars with
metallicities $10^{-2}, 0.025, 0.05$, and $0.1\zsun$ for Pop II stars,
and M-model stars with metallicities $10^{-8}, 10^{-6}$, and
$10^{-4}\zsun$ for Pop III and EMP stars as seen in Tables
\ref{tab:SingleStarAndPi} and \ref{tab:ParameterSets}. We do not use
SSE-model stars with $5 \times 10^{-3}\zsun$, despite that the SSE
model supports metallicities down to $5 \times 10^{-3} \zsun$. In
order to examine this, we prepare the SSE-plus set. In the SSE-plus
set, we add SSE-model stars with $5 \times 10^{-3}\zsun$ for Pop II
stars in addition to the metallicity list of the SSE-std set.

\begin{figure}[ht!]
  \plottwo{\fdir/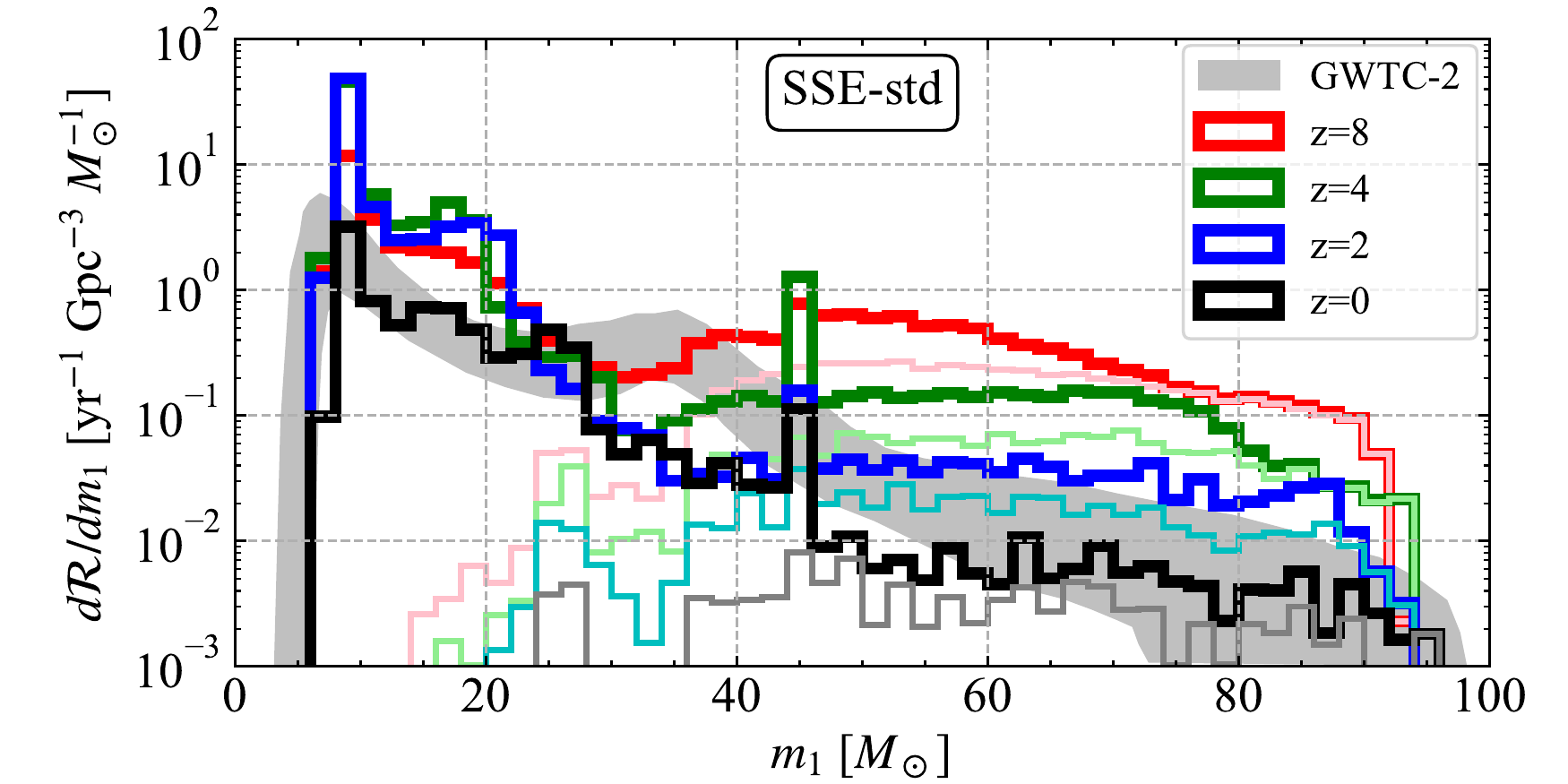}{\fdir/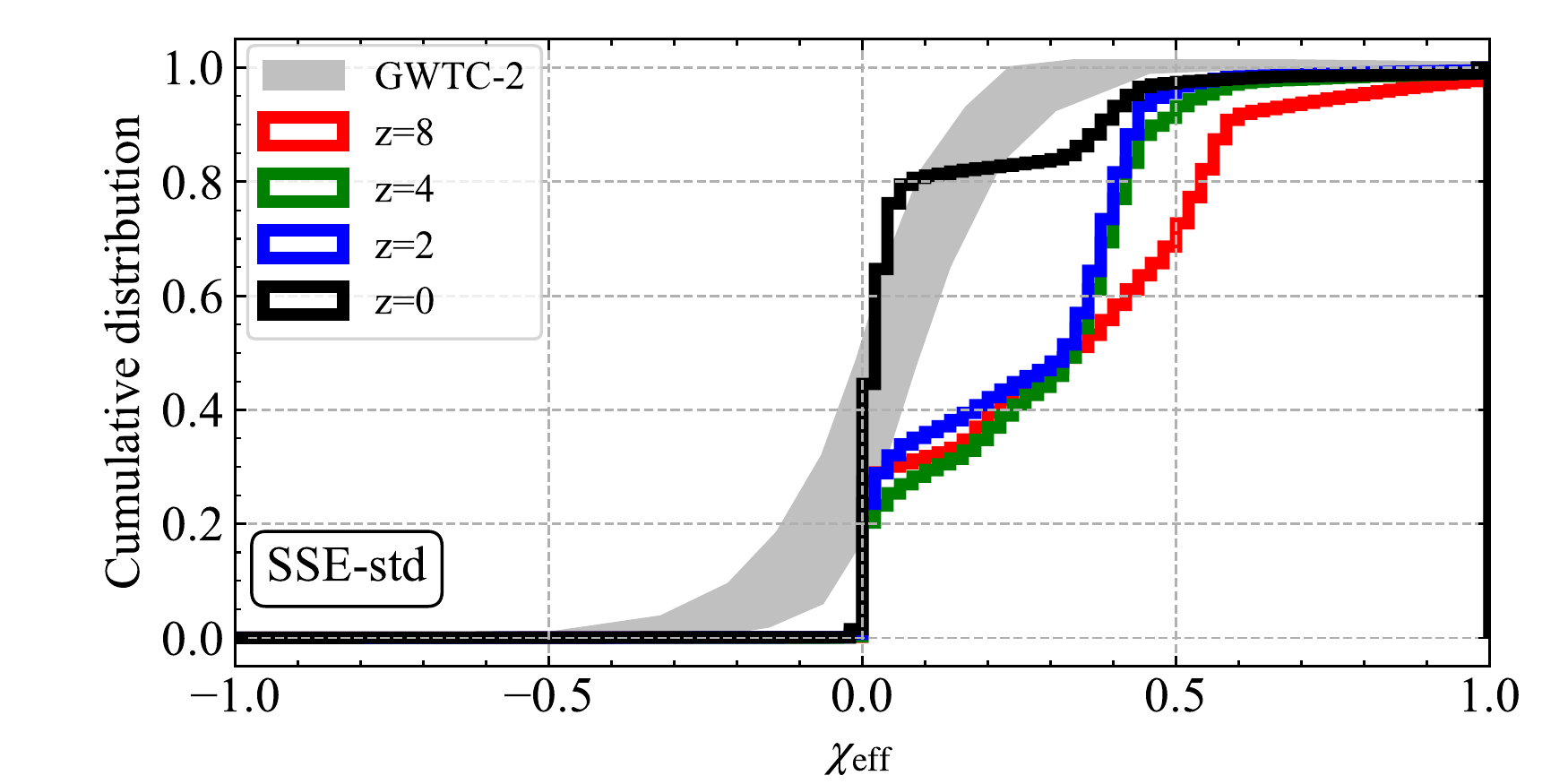}
  \plottwo{\fdir/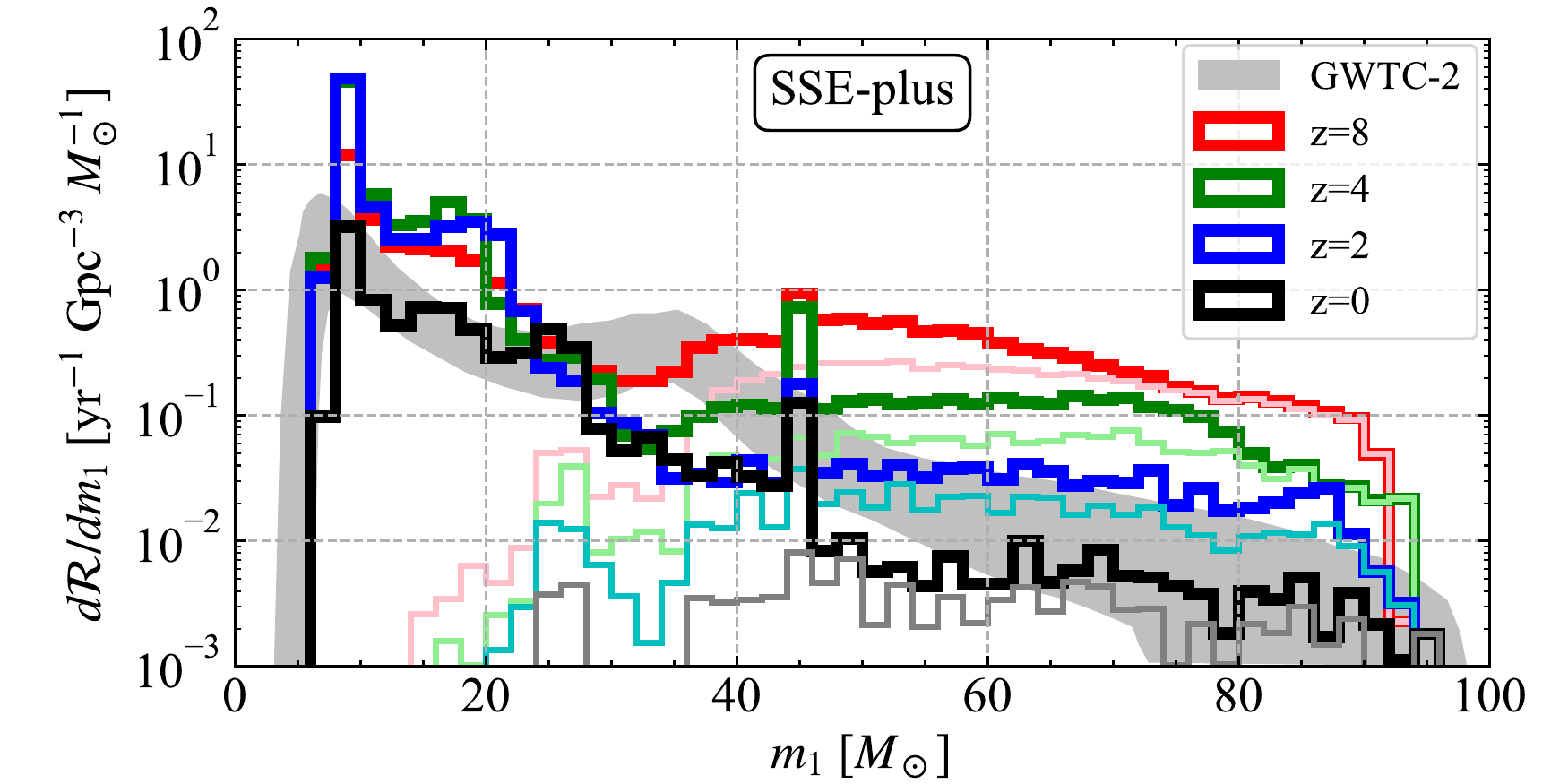}{\fdir/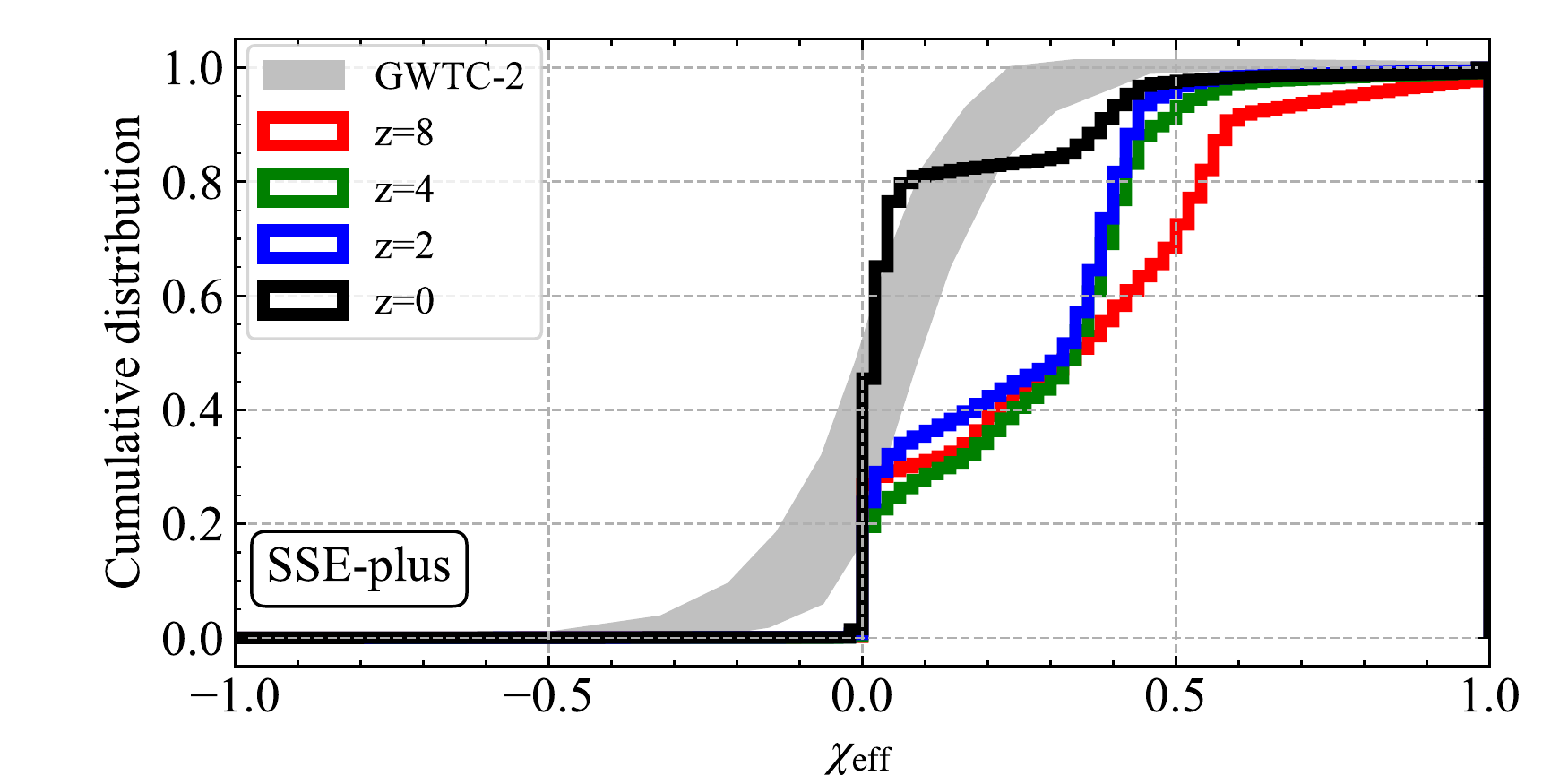}
  \caption{(Left) The same as Figure \ref{fig:mergerRateMassRedshift}
    except that results of the SSE-std and SSE-plus sets are shown in
    the top and bottom panels. (Right) The same as Figure
    \ref{fig:mergerRateXeffRedshift} except that results of the
    SSE-std and SSE-plus sets are shown in the top and bottom panels.}
  \label{fig:comparisonMetallicityChoice}
\end{figure}

Figure \ref{fig:comparisonMetallicityChoice} compares the mass and
spin distribution of binary BHs in the SSE-std and SSE-plus
sets. There is little difference between the two sets. Thus, we can
conclude that the not using SSE-model stars with $5 \times
10^{-3}\zsun$ has little effect on our results.

\end{document}